\documentclass[aps,prb,reprint,superscriptaddress,floatfix]{revtex4-2}

\usepackage{amssymb}
\usepackage{graphicx}   % need for figures
\usepackage{color}      % use if color is used in text
\usepackage{hyperref}   % use for hypertext links, including those to external documents and URLs
\usepackage{dcolumn}
\usepackage{todonotes}
\usepackage{amsmath}
\usepackage{orcidlink}
\usepackage[final]{changes}
\usepackage{booktabs}

\usepackage{glossaries}
\newacronym{NPD}{NPD}{neutron powder diffraction}
\newacronym{XPD}{XPD}{x-ray powder diffraction}
\newacronym{SCND}{SCND}{single crystal neutron diffraction}

\newacronym{TEM}{TEM}{transmission electron microscopy}
\newacronym{DFT}{DFT}{density functional theory}

\newacronym{FM}{FM}{ferromagnetic}
\newacronym{AFM}{AFM}{antiferromagnetic}
\newacronym{vdW}{vdW}{van der Waals}

\begin{document}

% \title{Distinct magnetic phases in Cr$_{3+\delta}$Te$_4$ crystals identified by single-crystal neutron diffraction}
%\title{Competing magnetic phases in Cr$_{3+\delta}$Te$_4$ do not coexist}
\title{Competing magnetic phases in Cr$_{3+\delta}$Te$_4$ are spatially segregated}

\author{V. K. Bhartiya\orcidlink{0000-0002-3575-7404}}
\email{vbhartiya1@bnl.gov}

\affiliation{Condensed Matter Physics and Materials Science Division, Brookhaven National Laboratory, Upton, New York 11973-5000, USA}
\author{Anirban Goswami\orcidlink{0009-0001-9800-2957}}
\affiliation{Department of Physics and Astronomy, Howard University, Washington, D.C. 20059, USA}
\author{Nicholas Ng}
\affiliation{Department of Chemistry, The Johns Hopkins University, Baltimore, Maryland 21218, USA}
\affiliation{Institute for Quantum Matter, The William H. Miller III Department of Physics and Astronomy,
The Johns Hopkins University, Baltimore, Maryland 21218, USA}
\author{Wei Tian\orcidlink{0000-0001-7735-3187}}
\author{Matthew G. Tucker\orcidlink{0000-0002-2891-7086}}
\affiliation{Neutron Scattering Division, Oak Ridge National Laboratory, Oak Ridge, Tennessee 37831, USA}
\author{Niraj Aryal\orcidlink{0000-0002-0968-6809}}
\author{Lijun Wu\orcidlink{0000-0002-8443-250X}}
\author{Weiguo Yin\orcidlink{0000-0002-4965-5329}}
\author{Yimei Zhu\orcidlink{0000-0002-1638-7217}}
\affiliation{Condensed Matter Physics and Materials Science Division, Brookhaven National Laboratory, Upton, New York 11973-5000, USA}
\author{Milinda Abeykoon\orcidlink{0000-0001-6965-3753}}
\affiliation{Photon Science Division, Brookhaven National Laboratory, Upton, New York 11973-5000, USA}
\author{Emmanuel Yakubu\orcidlink{0000-0002-7795-5775}}
\author{Samaresh Guchhait\orcidlink{0000-0002-0469-8034}}
\affiliation{Department of Physics and Astronomy, Howard University, Washington, D.C. 20059, USA}
\author{J. M. Tranquada \orcidlink{0000-0003-4984-8857}}
\email{jtran@bnl.gov}

\affiliation{Condensed Matter Physics and Materials Science Division, Brookhaven National Laboratory, Upton, New York 11973-5000, USA}

\date{\today} 

\begin{abstract}
Cr$_{1+x}$Te$_2$ is a self-intercalated \gls*{vdW} system that is of current interest for its room-temperature \gls*{FM} phases and tunable topological properties. In bulk samples, the strain from the interstitial Cr ions leads to distinct structural phases for different ranges of $x$.  Early \gls*{NPD} measurements on the monoclinic phase Cr$_3$Te$_4$ ($x=0.5$) presented evidence for competing \gls*{FM} and \gls*{AFM} phases.  Here we apply neutron diffraction to a single crystal of Cr$_{3+\delta}$Te$_4$ with $\delta=-0.10$ and discover that it consists of two distinct monoclinic phases, one with \gls*{FM} order below $T_{\rm C} \approx 321$~K and another that develops \gls*{AFM} order below $T_{\rm N} \approx 86$~K.  In contrast, we find that a crystal with $\delta=-0.26$ exhibits only \gls*{FM} order below $T_{\rm C} \approx 285$~K.  The single-crystal analysis is complemented by results obtained with \gls*{NPD}, \gls*{XPD}, and \gls*{TEM} measurements on the $\delta=-0.10$ composition.  From observations of spontaneous magnetostriction of opposite sign at $T_{\rm C}$ and $T_{\rm N}$, along with the \gls*{TEM} evidence for both monoclinic phases in a single thin ($\approx$ 100 nm) grain, we conclude that the two phases must have a fine-grained ($\lesssim$ 100 nm) intergrowth character, as might occur from high-temperature spinodal decomposition during the growth process.  Calculations of the relaxed lattice structures for the \gls*{FM} and \gls*{AFM} phases with \gls*{DFT} provide a rationalization of the observed spontaneous magnetostrictions.  Correlations between the magnitude and orientation of the magnetic moments with lattice parameter variation demonstrate that the magnetic orders are sensitive to strain, thus explaining why magnetic ordering temperatures and anisotropies can be different between bulk and thin-film samples, when the latter are subject to epitaxial strain.  Our results point to the need to investigate the supposed coexistence \gls*{FM} and \gls*{AFM} phases reported elsewhere in the Cr$_{1+x}$Te$_2$ system, such as in the Cr$_5$Te$_8$ phase ($x=0.25$).\\
\end{abstract}

\maketitle
\section{Introduction}

The family of compounds Cr$_{1+x}$Te$_2$ with $0\le x<1$ has long been of interest because of compositions with \gls*{FM} ordering above room temperature, and it has recently been the subject of renewed attention because of its potential for spintronics applications \cite{fuji20,wang22,mats24,chal24,guil24,kush25,ou25,he25}.  The end composition, CrTe$_2$, is a \gls*{vdW} compound composed of triangular layers of Cr and trigonal symmetry associated with the Te ligands.  On a coarse level, one can view the system as developing by intercalation of Cr into the CrTe$_2$ lattice \cite{fuji20}.  Detailed examination shows that the system breaks up into distinct structural families \cite{ipse83}, and these are often labeled by distinct chemical formulas representative of the different crystal structures.  In this paper, we will focus on the Cr$_{3+\delta}$Te$_4$ phase, which adopts a monoclinic structure for $-0.25\lesssim\delta\lesssim0.4$ (corresponding to $0.37\lesssim x\lesssim 0.7$) \cite{ipse83}.

Cr$_3$Te$_4$ was the subject of some of the earliest studies of magnetic order.  Powder neutron diffraction studies \cite{bert64,andr70} identified both a \gls*{FM} phase, with Curie temperature $T_{\rm C}\approx321$~K, and a coexisting \gls*{AFM} phase below N\'eel temperature $T_{\rm N} \approx 80$~K.  Evidence for the two phase transitions has also been detected in magnetization studies performed on single-crystal samples \cite{yama72,gosw24a}.  There were conflicting conclusions regarding the easy axis of the \gls*{FM} moments from the neutron and magnetization studies.  As we will show, \gls*{SCND} can help to resolve these issues; however, our results also provide some surprises.

\vspace{1.1cm}

To appreciate some of the complications involved, it is useful to consider the variations in  magnetic ordering that have been observed across the Cr$_{1+x}$Te$_2$ family. The first challenge involves the discontinuous evolution of the crystal structure.  The parent compound has a trigonal structure with lattice parameters $a=b=3.78$~\AA, $c=6.02$~\AA, and $\alpha=\beta=90^\circ$, $\gamma=120^\circ$ \cite{rose25},

\begin{figure}
    \centering
    \includegraphics[width=1\linewidth]{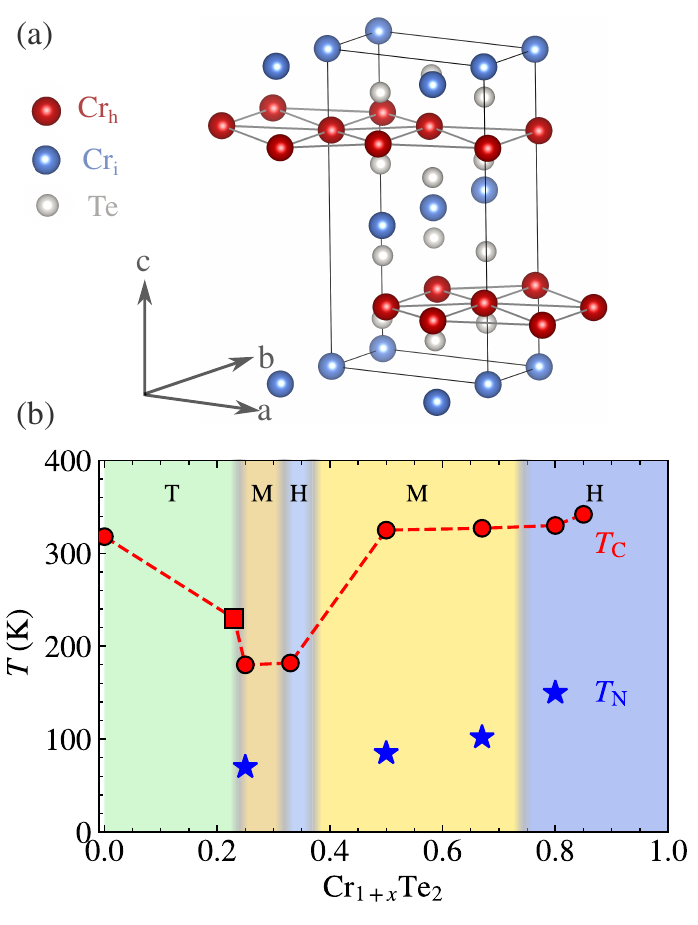}
    \caption{(a) The atomic arrangement in Cr$_3$Te$_4$. The hexagonal/pseudo-hexagonal Cr$_\text{h}$ layers (in magenta) are common to the Cr$_{1+x}$Te$_2$ system, while the occupied interstitial  Cr$_\text{i}$ sites (in blue) have a body-centered arrangement and vary among different phases. 
    (b) Approximate phase diagram for Cr$_{1+x}$Te$_2$. Phase boundaries are estimated from the data in Table II of Ipser {\it et al.} \cite{ipse83}. Letters denote symmetry: T:trigonal, M:monoclinic, and H:hexagonal. The circles and stars indicate the \gls*{FM} ($T_\text{C}$) and \gls*{AFM} ($T_\text{N}$) ordering temperatures, respectively, as listed in Table~\ref{tab}. The red square denotes $T_\text{C}$, which was measured by magnetization \cite{liu19}. }
    \label{fig:xtal_str}
\end{figure}

and it evolves to a similar cell in hexagonal Cr$_{1-\epsilon}$Te with $a=4.01$~\AA, $c=6.25$~\AA\ \cite{lotg57}. At intermediate values of $x$, the structure retains the 1T (or AA) stacking of the hexagonal Cr layers; however, it passes through various monoclinic and trigonal structures with enlarged unit cells due to the partial occupancy of interstitial sites \cite{ipse83}.  For convenience, we will label the Cr sites in the pseudo-hexagonal layers as Cr$_{\rm h}$ and the interstitial sites as Cr$_{\rm i}$, as indicated in Fig.~\ref{fig:xtal_str}(a).  

In Table \ref{tab}, we list the phases of Cr$_{1+x}$Te$_2$ that have been studied by \gls*{NPD}, with the commonly used chemical formulas and reported magnetic ordering temperatures; a graphical version of some of this information is presented in Fig.~\ref{fig:xtal_str}(b).  Note that the monoclinic structures of Cr$_5$Te$_8$ and Cr$_3$Te$_4$ are nearly orthorhombic, with $90^\circ<\beta<91.2^\circ$ \cite{ipse83}.

Given that our focus is on Cr$_3$Te$_4$, %which has a nearly-orthorhombic unit cell (in terms of space group $I2/m$),
for cell-size comparisons it is convenient to introduce an approximate unit cell based on an orthorhombic lattice and ignoring small distortions; converting from the trigonal/hexagonal unit cell with $a\sim4$~\AA\ and $c\sim6$~\AA, we make use of the orthohexagonal cell with $a'=\sqrt{3}a$, $b'= a$, and $c'= c$.  The unit cell and AFM cell comparisons in Table~\ref{tab} are based on this choice. 
 %convert from the trigonal/hexagonal unit cell to an orthorhombic cell with $a'=(\sqrt{3}/2)a$, $b'= a$, and $c'= c$ with $a\sim4$~\AA\ and $c\sim6$~\AA, and overlooking the small distortions of the cell for particular phases. 

The ordering of a fractional density of interstitials can lead to large unit cells.  Of course, when only a fraction of interstitial sites are filled, there is also the possibility of disorder from partial occupancy of the ``wrong" sublattice.  The large unit cells with two inequivalent sites allow for a variety of possible spin configurations.  The \gls*{FM} phases have typically been analyzed allowing for distinct magnetic moments on each sublattice but a common spin direction.  The \gls*{AFM} phases involve a considerably enlarged cell with additional moments and orientations.  The relationship between the \gls*{FM} and \gls*{AFM} phases has been unclear.  They appear to be associated with the same chemical phase; however, where the \gls*{FM} and \gls*{AFM} peak intensities have been plotted vs.\ temperature, as for Cr$_5$Te$_8$ \cite{huan08}, the development of the \gls*{AFM} order appears to have little impact on the \gls*{FM} order.  This is a bit surprising if the two orders are superimposed in the same lattice.

In this paper, we present a neutron diffraction study of nominal single crystals of Cr$_{3+\delta}$Te$_4$ with $\delta\approx -0.1$ and $-0.26$.  For the $\delta\approx-0.1$ crystal, measurements of rocking curves for a series of reflections indicate the presence of two dominant domains with distinct monoclinic angles and different magnetic orders: the larger one develops FM order below $\approx 321$~K, while the smaller one develops only AFM order below $\approx 86$~K.  In contrast, the $\delta\approx-0.26$ crystals exhibits only FM order.  We find supporting evidence for two different monoclinic phases in the $\delta\approx-0.1$ sample from \gls*{TEM}. To analyze the magnetic orders, we make use of \gls*{SCND} measurements at 4~K and \gls*{NPD} measurements (on ground crystals) at temperatures both below and above $T_{\rm N}$.  \gls*{XPD} data collected as a function of temperature provide evidence for substantial spontaneous magnetostriction that couples to the evolution of the magnitudes and orientations of magnetic moments.  %We end up with a picture of fine-scale spatial segregation of two phases with slightly different Cr content and very different magnetic orders, which can nevertheless interact through magnetostrictions of opposite sign.  This interpretation is supported by calculations of the optimal conditions for the FM and AFM phases with density functional theory.
These combined multimodal observations support a scenario of fine-scale spatial segregation of two phases with slightly different Cr content but distinct FM and AFM orders, which interact through magnetostrictions of opposite sign. This interpretation is supported by \gls*{DFT} calculations of the optimal conditions for the FM and AFM phases.  We note that it has been a challenge to fit together the pieces of this puzzle; further experiments that can confirm it are left for the future.

The rest of the paper is organized as follows. In the next section, we describe how the crystals were grown and characterized, along with details of the various diffraction measurements.  In Sec.~\ref{sec:RA}, we present and analyze the results in a step-by-step fashion to build a coherent understanding. In sec.~\ref{sec:SD}, we summarize our results and discuss their implications, followed by a brief conclusion.

\begin{table*}
\caption{Compositions of Cr$_{1+x}$Te$_2$ whose magnetic states have been investigated by \gls*{NPD} (and \gls*{SCND} in the case of Ohsawa {\it et al.} \cite{ohsa72}).  The unit cell is relative to the effective orthorhombic cell described in the text; the $x$ values are only approximate. Note that while $x=0.67$ corresponds to Cr$_5$Te$_6$ \cite{andr70}, its structure is the same as Cr$_3$Te$_4$ \cite{ipse83}. \label{tab}}
\begin{ruledtabular} 
\begin{tabular}{dcccdcccc} 
 x & Phase & space & unit & \multicolumn{1}{c}{Cr$_{\rm h}$/cell} & $T_{\rm C}$ & $T_{\rm N}$ & AFM cell & Refs. \\
  & & group & cell & & (K) & (K) & & \\
\hline
0\rule{0pt}{9pt} & CrTe$_2$ & $P\bar{3}m1$ & $1\times1\times1$ & 1 & 318 & \null & \null & \cite{rose25} \\ 
%0.25 & Cr$_5$Te$_8$-tr & $2\times2\times2$ & 8 & \null & \null & \null & \cite{ipse83} \\
0.25 & Cr$_5$Te$_8$ & $F2/m$ & $2\times2\times2$ & 16 & 180 & 70 & $2\sqrt{2}\times2\sqrt{2}\times2$ & \cite{huan08} \\
0.33 & Cr$_2$Te$_3$ & $P\bar{3}1c$ & $\sqrt{3}/2\times\sqrt{3}\times2$ & 6 & 182 & \null & \null & \cite{andr70,hama75} \\
0.5 & Cr$_3$Te$_4$ & $I2/m$ & $1\times1\times2$ & 4 & 329, 325 & 80, 85 & $2\times1\times4$ & \cite{bert64,andr70} \\
0.67 & Cr$_3$Te$_4$ & $I2/m$  & $1\times1\times2$ & 4 & 327 & 102 & $2\times1\times4$ & \cite{andr70} \\
1 & Cr$_{1-\epsilon}$Te & $P6_3/mmc$ & $1\times1\times1$ & 1 & 330, 342 & 150 & $1\times1\times1$ & \cite{cox65,take66,ohsa72} \\
\end{tabular}
\end{ruledtabular}
\end{table*}

\section{Experimental Methods}

%\subsection{Crystal growth and characterization}

%As described in \cite{gosw24a}, the crystals were grown by chemical vapor transport, starting with 99.99\%\ Cr and 99.9999\%\ Te and using I$_2$ as the transport agent.
Single crystals of Cr$_{3+\delta}$Te$_4$ were synthesized via the chemical vapor transport (CVT) method using high-purity elemental chromium and tellurium. A stoichiometric mixture of Cr powder (Alfa Aesar, $-100+325$ mesh, 99.99\%\ metals basis) and Te shots (Thermo Scientific, 2–5 mm diameter, 99.9999\%\ metals basis) was loaded into a fused quartz ampoule along with 30–50 mg of iodine (I$_2$) as a transport agent. The ampoule was sealed under vacuum and placed in a three-zone furnace (Thermo Scientific Lindberg Blue M) equipped with UP150 program controllers. The CVT process was initiated by ramping the temperature at a rate of 100 $^\circ$C/h, reaching 1055 $^\circ$C in the charge zone and maintaining 820 $^\circ$C in the growth zone. This temperature gradient was held constant for 8 days to facilitate crystal growth. The furnace was then cooled to room temperature at the same rate of 100 $^\circ$C. The resulting crystals appeared gray in color and had typical dimensions of 2–5 mm. Closer examination indicated two types of crystals with a subtle distinction in color.  Magnetization measurements confirmed different $T_{\rm C}$ values of $\approx321$~K \cite{gosw24a} and $\approx285$~K.

To confirm the phase, structure, and composition, we performed single-crystal x-ray diffraction %(SCXRD) 
measurements on small crystals, approximately $0.05\times0.03\times0.01$~mm$^3$, at 293(2)~K using a Rigaku SuperNova diffractometer equipped with an Atlas detector and using Mo K$\alpha$ radiation ($\lambda = 0.71073$~\AA); the temperature was regulated using an Oxford Instruments Cryojet system. Data collection and reduction were performed using CrysAlisPro (Version 1.171.42.49, Rigaku OD, 2022). The crystal structure was solved using SHELXS-2018/2 and refined on $F^2$ using SHELXL-2019/3. An analytical numeric absorption correction was applied using a multifaceted crystal model \cite{clar95} within CrysAlisPro. The fits to the x-ray data indicated $\delta=-0.10$ for the first crystal and $\delta=-0.26$ for the second, corresponding to Cr$_{2.90}$Te$_4$ with the higher $T_{\rm C}$ and Cr$_{2.74}$Te$_4$.

Some crystals of the $\delta=-0.10$ growth were ground to powder and measured by \gls*{XPD} at the PDF beamline, 28-ID-1, of the National Synchrotron Light Source II, Brookhaven National Laboratory \cite{gosw24a}.  Those data were analyzed by Rietveld refinement.  As an example, a room-temperature measurement is reported in \cite{gosw24a}.  When the results are converted from space group $C2/m$ to $I2/m$, the unit cell parameters are comparable to those of the single-crystal characterization.  Data were also collected and analyzed at $T=34$, 200, and 349~K. The atomic coordinates change little with temperature; the main change is in the unit-cell parameters.
 %using the standard setting of space group no.~12, $C2/m$, where, at $T=34$~K, the unit cell is characterized by $a=13.921$~\AA, $b=3.934$~\AA, $c=6.868$~\AA, and $\beta=118.36^\circ$.  In order to have a nearly orthorhombic unit cell, we will work with the setting $I2/m$, for which the corresponding parameters are $a=6.868$~\AA, $b=3.934$~\AA, $c=12.31$~\AA, and $\beta=91.1^\circ$.

A second powder sample was studied by \gls*{NPD} on the POWGEN instrument at the Spallation Neutron Source, Oak Ridge National Laboratory (ORNL).  The sample was loaded into a cylindrical vanadium can mounted in a cryostat and data were collected at $T=5$, 80, 120, 300, 320, 350, and 400 K, using a center wavelength of 1.5~\AA\ and yielding data in the momentum-transfer range of $Q = 0.7$--12.5~\AA$^{-1}$. To obtain a complementary perspective, a small powder grain was studied by \gls*{TEM} using a 200 kV beam, corresponding to a wavelength of 0.0251~\AA. 

\begin{figure*}[t]
 \centering
 \includegraphics[width=2\columnwidth]{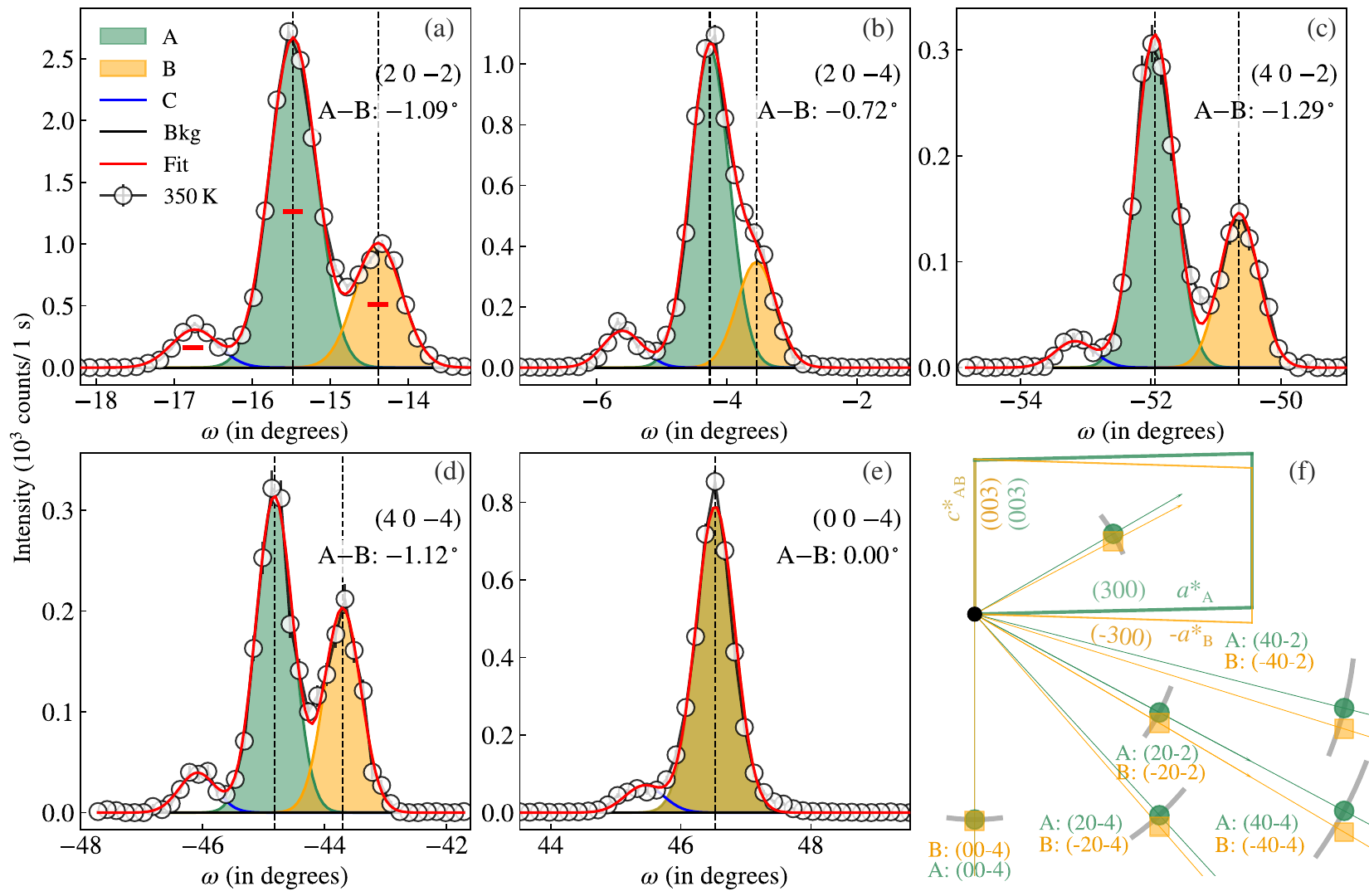}
    \caption{\label{fg:rock} (a-e) Rocking curves at 350 K for a selection of structural Bragg peaks illustrating the varying separation of peaks A and B. Legends are common across all the panels. The calculated instrument resolution is represented by the horizontal bars (red) in (a). The fit (red line) to rocking curves consists of three peaks: A (seagreen), B (orange), and C (blue), simulated by Gaussian lineshapes of fixed width, and a global linear background (black). The vertical dashed lines show the position of A and B peaks, and their difference is highlighted in the panels by A$-$B. (f) A schematic highlighting the orientation of phase B with respect to phase A in the ($H$ 0 $L$) reciprocal space. Their $c^*$ axes are co-aligned %, $c^*_\text{A}$ and $c^*_\text{B}$ points in the same direction- 
    and represented by $c^*_\text{AB}$.  The $a^*$ axis associated with the B peaks is rotated by 180$^\circ$ around the $c^*$ axis relative to the $a^*$ axis of the A peaks, such that $a^*_\text{A}$ and $- a^*_\text{B}$ are aligned, and their angular difference is equal to $\beta_{\rm A}+\beta_{\rm B}-180^\circ$.}
\end{figure*}

To study the single crystals, \gls*{SCND} measurements were performed on the VERITAS (HB-1A) triple-axis spectrometer at the High Flux Isotope Reactor, ORNL.  For the monochromator and analyzer, we used the (0 0 2) reflection of pyrolytic graphite, with the fixed incident energy of 14.5 meV; horizontal collimations were set to $40'$-$40'$-S-$40'$-$80'$. The $\delta=-0.10$ crystal, with a mass of 23.7 mg, was wrapped in Al foil and mounted on a thin Al disk clamped on a sample base attached to the sample stick of a Janus top-loading cryostat.   We aligned it so that ($H$ 0 $L$) reflections were in the horizontal scattering plane.  The crystal of $\delta=-0.26$ with a mass of 17 mg was studied in a similar fashion in a separate experiment.

\section{Results and Analysis}
\label{sec:RA}

\subsection{Evidence for two phases in $\delta$ = $-0.1$}

%From rocking curves, we identify phases A and B, which have different values of $\beta$; also, they are aligned along $c$ but have a relative rotation by 180$^\circ$ of their $a$ axes.  See Fig.~\ref{fg:rock}.

When we first aligned the crystal at room temperature, it was difficult to detect the (2 0 0) and (4 0 0) peaks (because their nuclear intensities are negligible), so we worked with the (0 0 $-8$), (2 0 2), and (2 0 $-2$).  After mounting in the cryostat and cooling to 4~K, we were able to align on the (2 0 0) and the (0 0 $-8$).  From the orientation matrix, we obtained $a=6.815$~\AA, $c=12.216$~\AA, and $\beta=90.35^\circ$. [We could not measure $b$ from the crystal, as we only had access to the $(H0L)$ plane; however, analysis of the \gls*{NPD} measurements, discussed shortly in Sec.~\ref{sec:snpd}, yielded $b=3.93$~\AA].  

In Fig.~\ref{fg:rock}, we show a set of rocking curves (rotating the crystal with the spectrometer at fixed wave vector {\bf Q}) displaying multiple peaks for several different reflections. If there were multiple crystal domains of the same phase but with different orientations, we would expect the same set of peaks to show up with identical angular separations for all reflections.  Instead, we find that the peaks labeled A and B have a variable angular separation; in fact, we cannot resolve separate peaks for (0 0 $-4$).
Note that there is also a third peak C; however, we could not simultaneously align the ${\bf a}^*$ and ${\bf c}^*$ axes of all three sets of peaks within the scattering plane, and so we will not do any quantitative analysis of C.

We had initially aligned the sample using the A peaks resulting in $a_{\rm A}=6.815$~\AA\,and $\beta_{\rm A}=90.35^\circ$.  When we tried aligning on peak B for the (4 0 0) reflection, we found $a_{\rm B}=6.877$~\AA\ and $\beta_{\rm B}=89.22^\circ$. This immediately leads to two conclusions. First of all, it indicates that peaks A and B correspond to distinct phases.  Secondly, we can recover a value of $\beta\ge90^\circ$ for phase B if we view it as rotated by $180^\circ$ around ${\bf c}^*$ relative to phase A.  In this case, (4 0 0)$_{\rm A}$ is adjacent to ($-4$ 0 0)$_{\rm B}$ and $\beta_{\rm B}=90.78^\circ$.  The mosaic of each of these phases is broader than the calculated instrument resolution indicated by the horizontal bars in Fig.~\ref{fg:rock}(a). The correspondence between the reflections for phases A and B is illustrated in Fig.~\ref{fg:rock}(f).    Properly distinguishing these phases turns out to be important, as they exhibit different magnetic orders, as we will discuss shortly.

Given two phases with slightly different lattice parameters, one might expect to see a splitting when measuring along {\bf Q} in a $\theta$-$2\theta$ scan; however, we observed no evidence of any peak splittings or notable peak broadenings.  Similarly, the neutron powder diffraction data, to be discussed, did not give direct evidence for two different structural phases.  The only indication of two phases came from the rocking-curve scans.

To appreciate the relative orientations of phases A and B, we note that ${\bf c}^*$ is defined to be perpendicular to the real-space vectors {\bf a} and {\bf b}.  The fact that the ${\bf c}^*$ vectors of phases A and B are parallel means that the pseudo-hexagonal planes of these phases are parallel, while the orientation of {\bf a} reverses in moving from one phase to the next. 

\subsection{Distinguishing the magnetic orders of phases A and B}

\begin{figure}[t]
 \centering
    \includegraphics[width=0.9\columnwidth]{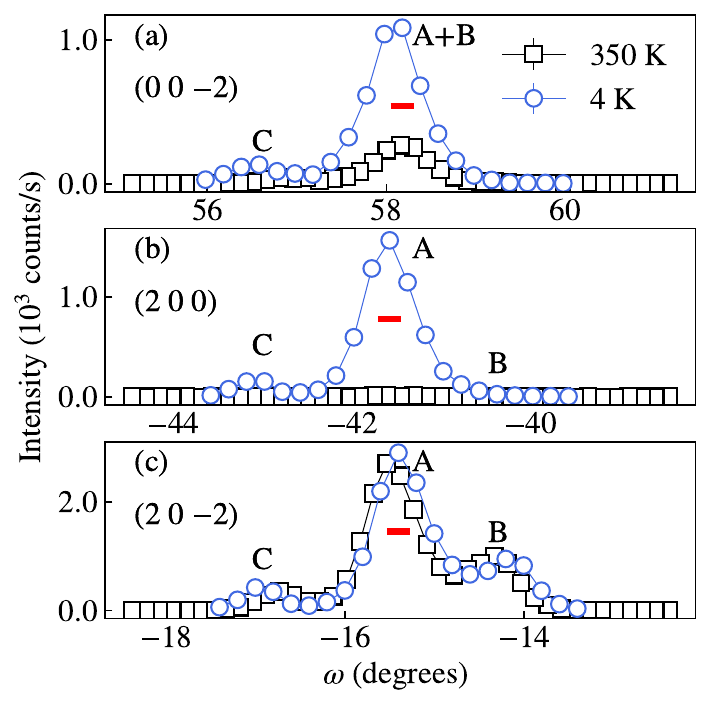}
    \caption{\label{fg:200} Rocking curves at 4 K (circles) and 350 K (squares) for (0 0 $-2$), (2 0 0), and (2 0 $-2$). Letters A, B, and C highlight the locations of peaks from the respective phases. The calculated instrument resolution is represented by the horizontal bars (red). }
\end{figure}

In Fig.~\ref{fg:200}, we compare rocking curves for several reflections measured at $T=4$ and 350~K.  The (2 0 0) and (0 0 $-2$) peaks of phase A show a large $T$ dependence at nuclear positions, consistent with \gls*{FM} order.  In contrast, the $T$ dependence of the B peaks is negligible, indicating an absence of ferromagnetism in that phase. The widths of the magnetic (2 0 0) and (0 0 $-2$) peaks of phase A are broader than the calculated instrument resolution (horizontal bars in Fig.~\ref{fg:200}) but similar to that of the (2 0 $-2$) nuclear peak, suggesting that the widths are limited by the sample mosaic.

\begin{figure}[t]
 \centering
    \includegraphics[width=0.9\columnwidth]{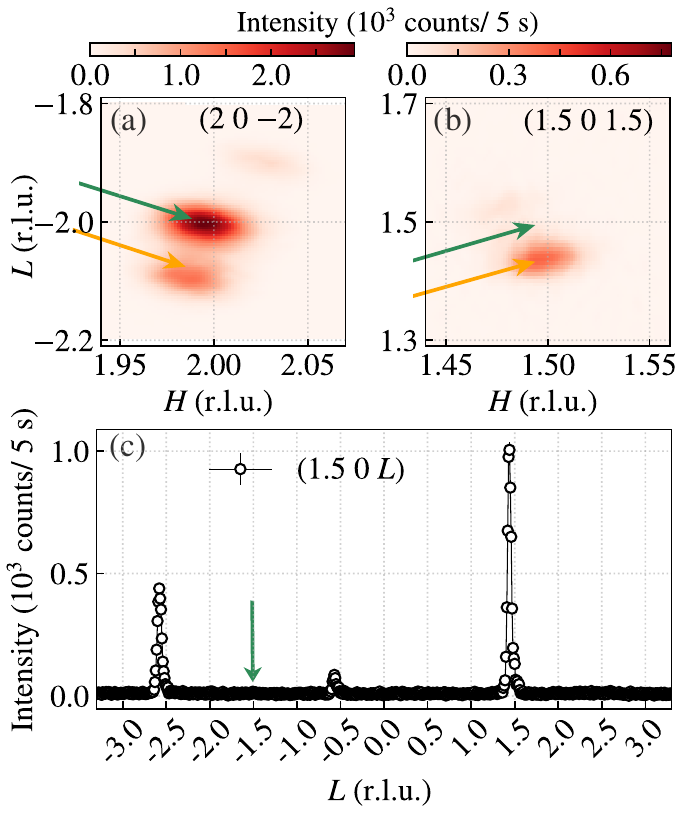}
    \caption{\label{fg:mesh} Mesh scans for (a) (2 0 $-2$) and (b) (1.5 0 1.5) based on the indexing of phase A. (c) Scan along ${\bf Q}$ = (1.5 0 $L$)$_{\rm A}$; green arrow points to the absence of a peak for phase A at (1.5 0 $-1.5$). }
\end{figure}

The AFM peaks identified in previous \gls*{NPD} studies \cite{bert64,andr70} appear at ${\bf Q}=(\frac{2m+1}{2}\, 0\, \frac{2n+1}{2})$ for integer $m,n$ with the constraint that $m+n$ is odd.  Figure~\ref{fg:mesh}(b) shows a mesh scan for (1.5 0 1.5) indexed on phase A, which does not satisfy the constraint; the finite peak corresponds to the phase B lattice, with indexing ($-1.5$ 0 1.5), which does.  To determine whether phase A exhibits AFM order, consider the scan along (1.5 0 $L$) in Fig.~\ref{fg:mesh}(c).  One can clearly see that there is no peak for $L=-1.5$; therefore, we conclude that phase A has FM order but no AFM order.

To determine the magnetic moments associated with these distinct phases, we need to be able to normalize the magnetic intensities to the nuclear intensities.  Since we only measured diffraction peaks in one plane of reciprocal space for the crystal and cannot distinguish the contributions to (0 0 $L$) reflections, we first consider analysis of the structure and effective magnetic moments from the \gls*{NPD} data.

\subsection{Structure from \gls*{NPD}}
\label{sec:snpd}
% 350 K data
\begin{table*}
\caption{Results of the GSAS analysis of the \gls*{NPD} data for the $\delta=-0.10$ sample at $T=350$~K using space group $I2/m$; the refined unit cell parameters are $a =  6.8543(1)$~\AA, $b = 3.9268(1)$~\AA, $c = 12.3092(3)$~\AA, and $\beta= 91.053(2)^\circ$.  \label{tb:struc350}}
\begin{ruledtabular}
\begin{tabular}{ccccccc}
% \hline
%\hline
 Atom &  site &  x &  y &  z &  U (\AA$^2$) & \multicolumn{1}{c}{ Occupancy} \\
\hline
% \multicolumn{7}{c}{ 350 K} \\
% \hline
Cr$_{\rm h}$ & 4$i$ & 0.9815(4) & 0.00000 & 0.2544(3) & 0.0057(5) & 1 \\
Cr$_{\rm i}$ & 2$a$ & 0 & 0 & 0 & 0.0075(8) & 0.90 \\
Te & 4$i$ & 0.6643(3) & 1.00000 & 0.6327(2) & 0.0040(4) & 1 \\
Te & 4$i$ & 0.8353(3) & 0.50000 & 0.3810(2) & 0.0056(4) & 1 \\
%\hline
\end{tabular}
\end{ruledtabular}
\end{table*}

\begin{figure}[b]
    \centering
    \includegraphics[width=0.9\linewidth]{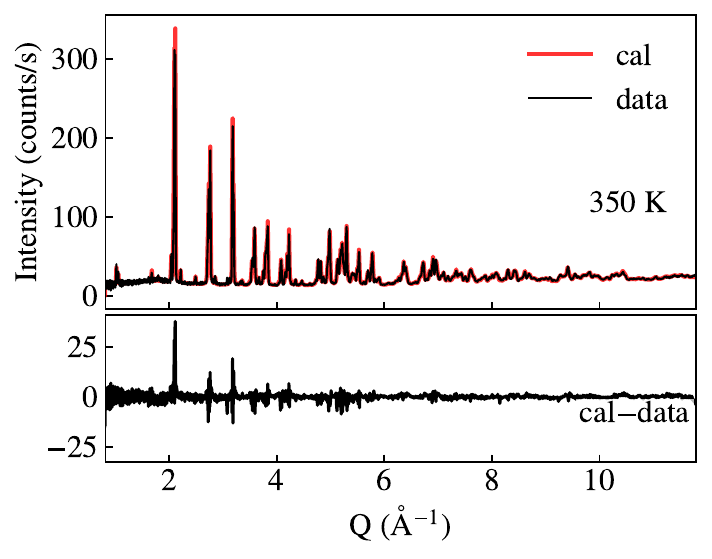}
    \caption{Comparison of the the structural refinement (red line) from the 350 K \gls*{NPD} data (black line) (the weighted-profile factor $R_\text{wp} = 5.05$) for the $\delta = -0.10$ sample. Lower panel shows the difference between calculated intensities and the data.}
    \label{fig:NPD_fit_350K}
\end{figure}

\begin{figure*}[t]
 \centering
    \includegraphics[width=1.9\columnwidth]{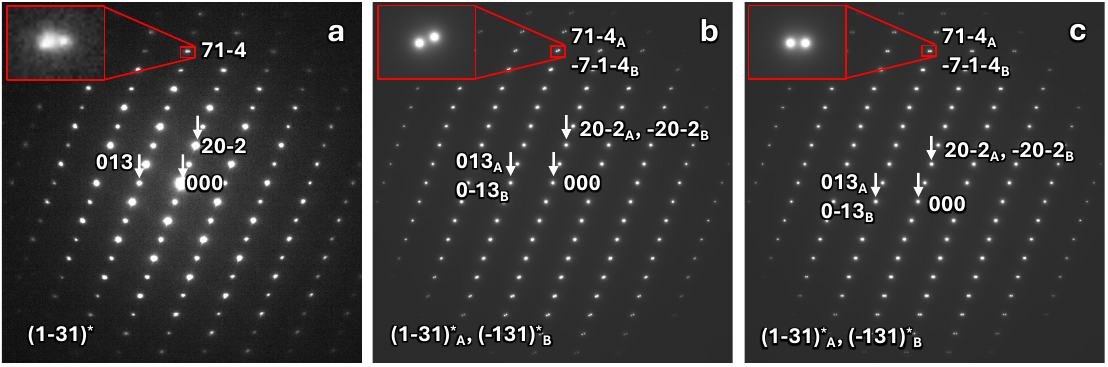}
    \caption{\label{fg:tem}  (a) Electron diffraction pattern (EDP) with a $[1,-3,1]$ zone axis for a grain of the $\delta=-0.10$ sample. (b) Simulated EDP for two domains with the same lattice parameters ($a=6.85$~\AA) but rotating the lattice of domain B 180$^\circ$ around the ${\bf c}^*$ axis. (c) The same as (b) but changing the $a$ lattice parameter for domain B from 6.85 to 6.90~\AA.}
\end{figure*}

The crystal structure (and the magnetic structures, as discussed later) were refined from \gls*{NPD} using the GSAS-II software suite \cite{Toby_GSAS_2013}. The starting model was  based on the $I2/m$ structure \cite{bert64}, with Wyckoff sites indicated in Table~\ref{tb:struc350}.  All atomic coordinates, isotropic displacement parameters, lattice parameters, and the scale factor were refined.  As in previous \gls*{NPD} studies of $\delta\approx 0$ samples \cite{bert64,andr70}, both the FM and AFM phases were detected, which, based on our single-crystal results, indicates the presence of two compositions.  Attempts to allow for the presence of two structural phases did not yield reliable results, so only a single phase was used for the refinement. An anisotropic strain tensor was refined to fully account details in the lineshapes, which is likely necessary due to the presence of two phases.  (High-resolution \gls*{XPD} would be necessary to properly resolve the structural phases.) For the paramagnetic phase at 350~K, the data and fit are shown in Fig.~\ref{fig:NPD_fit_350K} and the resulting structural parameters are listed in Table~\ref{tb:struc350}.

%A \gls*{FM} ($\mathbf{k} = (0\, 0\, 0)$ in $I2^\prime/m^\prime$ space group) and \gls*{AFM} ($\mathbf{k} = (\frac{1}{2}\, 0\,0)$ in $c2/m$ space group) phase were introduced below their respective ordering temperature. The \gls*{FM} magnetic space group constraints spins in the $ac-$ plane and \gls*{AFM} along the $b-$ axis. These are the only possible space groups since the magnetic moment plane (\gls*{FM}) and direction (\gls*{AFM}) has been determined independently with \gls*{SCND}. The magnetic moment magnitudes at two Cr sites were the only magnetic parameter refined. The \gls*{FM} moments at 5 K for Cr$_\text{h}$ [mx, my, mz] = [1.857(0.050) 0 $-$1.674(0.064)] $\mu_\text{B}$ and for Cr$_\text{i}$ = [1.971 (0.095) 0 $-$1.103(0.103)] $\mu_\text{B}$. For \gls*{AFM} phase the fit does not work well (R-factor is 20$\%$), producing Cr$_\text{h}$ (mx, my, mz) = [0 0.154(0.163) 0] $\mu_\text{B}$ and for Cr$_\text{i}$ = [0 -0.356(0.106) 0] $\mu_\text{B}$ small magnetic moment with large error bar. A few representative fitted \gls*{NPD} datasets are shown in Fig. and  the fitted lattice parameters from sequential fitting over all measured temperature are presented in Fig. \ref{fg:latt}(a). The goodness of global fit metrics for all temperatures were below R$_w$ $<$ 7$\%$.

\subsection{Evidence for two phases from TEM}

While the \gls*{SCND} reveals direct evidence for the presence of two distinct phases, they are not readily resolved in the powder diffraction data.  Hence, the question of the relationship between these structural phases remains open.  Are these macroscopic domains, such that both domains would be unlikely to appear in a small grain?  To test this, electron diffraction patterns (EDPs) were measured on a grain with a thickness of order 100~nm. 

An example of an EDP with a [1, $-3$, 1] zone axis is shown in Fig.~\ref{fg:tem}(a).  An enlargement of the (7 1 $-4$) reflection shows a splitting into two peaks.  In a first attempt to simulate this behavior, Fig.~\ref{fg:tem}(b) shows the case of two domains, both with the same $a$ lattice parameter, but with one domain B rotated $180^\circ$ around ${\bf c}^*$, so that the reflection of the second domain corresponds to ($-7$ $-1$ $-4$).  The resulting pair of peaks have an appropriate splitting, but the alignment is rotated.  To simulate the correct alignment, Fig.~\ref{fg:tem}(c) shows the case in which the rotated domain also has its $a$ lattice parameter increased from 6.85 to 6.90~\AA.  This gives reasonable agreement with the measured EDP and is also consistent with the single-crystal neutron diffraction results.

\subsection{Normalization of \gls*{SCND}}

In order to evaluate the magnetic order, we first need to determine the volumes of the two phases, A and B, in the sample.  To do this, we analyze the observed nuclear peak intensities at 350~K, where no magnetic order is present.  Our measurement geometry limits us to ($H$ 0 $L$) reflections, and the overlap of A and B phases for (0 0 $L$) further restricts the number of usable peaks.  Given the limited number of independent reflections, we chose to simply fit a scale factor to the data for each phase, using structural parameters from the powder neutron diffraction analysis at 350~K in Table~\ref{tb:struc350}.

\begin{figure}[t]
\includegraphics[width=0.9\columnwidth]{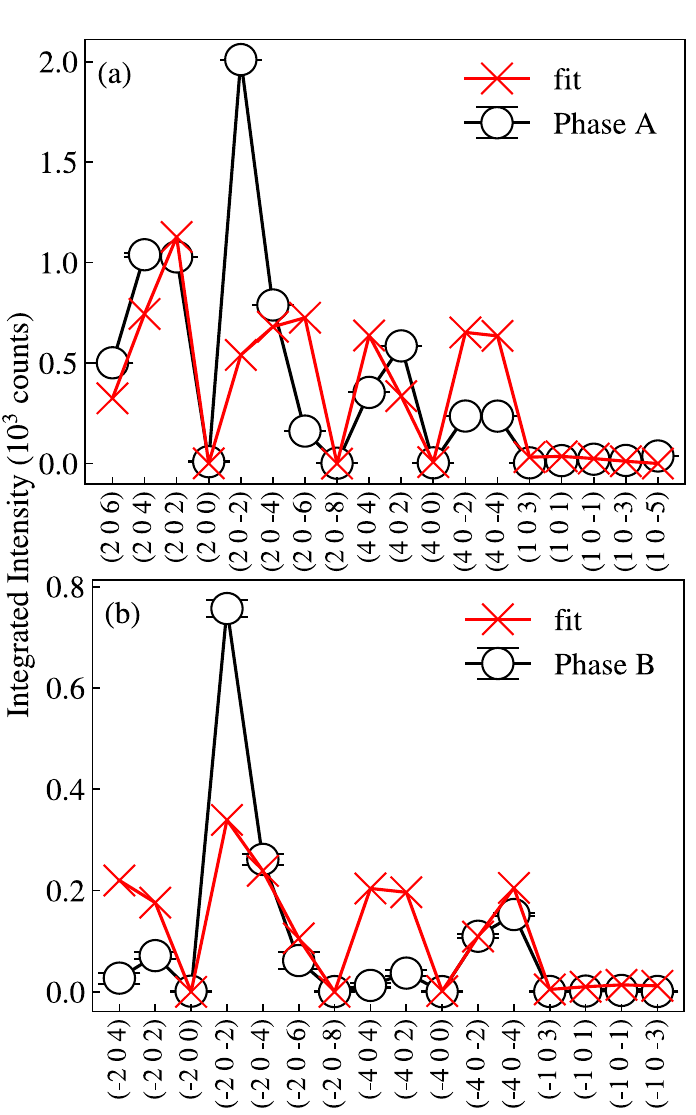}
\caption{\label{fg:nucl} Comparison of measured integrated intensities (circles) and fitted results (crosses) for phases A (upper panel) and B (lower panel) at 350 K, where scale factor is the only fitting parameter.}
\end{figure}

A complication concerns the occupancy of the interstitial Cr site in each phase. The fact that we have two phases with distinct magnetic orders within the sample suggests that we have two slightly different compositions present.  Such a result would seem to be consistent with the complicated thermal phase diagram for Cr-Te \cite{ipse83}.  As we will discuss later, stability of the AFM phase, with its large unit cell, would likely be favored by a stoichiometric composition, whereas the FM order is observed over a wide composition range (including our $\delta=-0.26$ sample). Hence, in applying the powder diffraction results, we choose to set the occupancy for the interstitial Cr site to 1 for phase B and to 0.9 for phase A.  In any case, these choices have minimal impact on the fits.  A comparison of the measured and fitted intensities for phases A and B are shown in Fig.~\ref{fg:nucl}.  From the respective fits, we find that the total volume is 75.7\%\ phase A and 24.3\%\ phase B.

\subsection{Analysis of magnetic symmetry}
\label{sec:ms}

We used the Bilbao Crystallographic Server (BCS) to analyze the irreducible representations (Irreps) associated with both the FM and AFM ordering. The previous neutron diffraction studies \cite{bert64,andr70} indexed the \gls*{FM} and \gls*{AFM} phases in terms of the $I2/m$ cell with propagation vector $[0 0 0]$ and  $[\frac12\, 0\, -\frac12]$, respectively. The space-group symbol \textit{I2/m} does not appear in BCS as a separate entry because it is a non-standard setting of the standard monoclinic space group \textit{C2/m} (No.~12) defined in the \textit{International Tables for Crystallography} (Vol.~A). Therefore, all symmetry analysis for both FM and AFM cases was carried out in the standard \textit{C2/m} setting. The FM and AFM propagation vectors translates to $[000]$ and $[00\frac12]$ in the $C2/m$ setting, respectively.  The Irreps result are listed in Table~\ref{tab:symmetry_analysis}.

\begin{table}[h]
\caption{Magnetic subgroups, associated Irreps, and magnetic moment constraints for Cr$_\text{h}$ and Cr$_\text{i}$ sites (Wyckoff positions 4i and 2a, respectively) for magnetic propagation vectors $\mathbf{k} = [000]$ and $\mathbf{k} = [00\frac12]$, based on the $C2/m$ lattice symmetry of Cr$_3$Te$_4$.}
\label{tab:symmetry_analysis}
%\centering
\begin{ruledtabular}
\begin{tabular}{lllcc}
%\toprule
\multicolumn{1}{c}{$\mathbf{k}$} & \multicolumn{1}{c}{Subgroup} & Irrep  & 4i(Cr$_\text{h}$) & 2a(Cr$_\text{i}$) \\
%\midrule
\hline
$[000]$ & $C2'/m'$ & mGM$_2^+$ & $(m_x, 0, m_z)$ & $(m_x, 0, m_z)$ \\
        & $C2/m'$  & mGM$_1^-$ & $(m_x, 0, m_z)$ & $(0, 0, 0)$     \\
        & $C2'/m$  & mGM$_2^-$ & $(0, m_y, 0)$   & $(0, 0, 0)$     \\
        & $C2/m$   & mGM$_1^+$ & $(0, m_y, 0)$   & $(0, m_y, 0)$   \\
%\midrule
\hline
$[00\frac12]$ & $C_c2/c$ & mA$_2^+$ & $(m_x, 0, m_z)$ & $(m_x, 0, m_z)$ \\
              & $C_c2/c$ & mA$_1^-$ & $(m_x, 0, m_z)$ & $(0, 0, 0)$     \\
              & $C_c2/m$ & mA$_2^-$ & $(0, m_y, 0)$   & $(0, 0, 0)$     \\
              & $C_c2/m$ & mA$_1^+$ & $(0, m_y, 0)$   & $(0, m_y, 0)$   \\
\bottomrule
%\hline
%\hline
\end{tabular}
\end{ruledtabular}
\end{table}

For the FM case, described by ordering wave vector $\mathbf{k} = [000]$, %two $k$-maximal magnetic subgroups are allowed: $C2/m$ and $C2'/m'$. The subgroup $C2/m$, associated with the Irrep mGM$_{1}^{+}$, restricts the magnetic moments to the $b$-axis. In contrast, $C2'/m'$, corresponding to the Irrep mGM$_{2}^{+}$, allows spins to lie within the $a$-$c$ plane. 
there are 4 Irreps, of which two restrict the magnetic moments to the $b$-axis and the other two constrain the spins to the $a$-$c$ plane.  For each of these pairs of Irreps, one allows finite moments on both Cr$_{\rm h}$ and Cr$_{\rm i}$ sites while the other has zero moment on Cr$_{\rm i}$. As will become clear in Sec.~\ref{sec:FM}, the first FM Irrep, mGM$_2^+$, with moments in the $a$-$c$ plane for both Cr sites, is the one consistent with the single-crystal diffraction data.  We will do the analysis in terms of the $I2/m$ unit cell, but this does not create any complications in applying the moment constraints, as the $a$-$c$ planes of the $I2/m$ and $C2/m$ settings are parallel with one another.

For the AFM case with propagation vector $\mathbf{k} = [00\frac{1}{2}]$, %two $k$-maximal magnetic subgroups are allowed: $C_c2/c$ and $C_c2/m$. The subgroup $C_c2/c$ restricts magnetic moments to the $a$-$c$ plane and corresponds to the Irrep mA$_{2}^{+}$, while $C_c2/m$ corresponds to a single Irrep mA$_{1}^{+}$ and allows moments along the $b$-axis only. 
we have another 4 Irreps, again with moments either along $b$ or in the $a$-$c$ plane, and with moments on both sites or only on Cr$_{\rm h}$.
We will see in Sec.~\ref{sec:AFM} that the Irrep mA$_1^+$, with moments on both sites and constrained to the $b$ axis, provides a good fit to the single-crystal diffraction results. %Therefore, the AFM refinement was performed using $C_c2/m$.
We will show the data indexed in the $I2/m$ setting, which involves a magnetic cell volume that is quadrupled compared to the nuclear cell; however, as we will illustrate, the $C_c2/m$ setting provides a magnetic cell with just double the volume of the nuclear cell.

%We used ISODISTORT \cite{Campbell_2006} to identify the appropriate magnetic space groups for the FM and AFM phases. For the \gls*{FM} phase, with propagation wave vector  $\mathbf{k} = [0\, 0\, 0]$, the only option is the $I2^\prime/m^\prime$ magnetic space group, with the constraint that the spins lie within the $a$-$c$ plane.  

%For the \gls*{AFM} order, to describe the only option for magnetic space group it is necessary to change from the $I2/m$ cell choice in space group 12 to $C2/m$, which reversed the direction of {\bf b} while putting the c-axis (${\bf c}'$) parallel to the pseudo-hexagonal layers and changing from $\beta  = 90^\circ$ to $\beta' =120^\circ$.  Within the $C2/m$ cell, the magnetic space group is $C_c2/m$, with propagation vector $\mathbf{k} = [0\, 0\, \frac{1}{2}]$; the spins are constrained to lie along $[0,1,0]$.

%  With the moments constrained to be along {\bf b}, this is an equivalent order but involves a magnetic unit cell of twice the volume required for $C_c2/m$.  We will show both options when we consider the \gls*{AFM} order.

\subsection{FM phase from \gls*{SCND} for $\delta=-0.1$}
\label{sec:FM}

An issue that has been the subject of considerable discussion concerns the orientation of the FM moments.  Neutron diffraction is sensitive to the spin direction, as the magnetic intensity for a particular wave vector {\bf Q} is proportional to the square of the spin component perpendicular to {\bf Q}.  In their \gls*{NPD} studies, Bertaut {\it et al.} \cite{bert64} and Andresen \cite{andr70} reported that the spins are along ${\bf a}$; however, this was not a unique fit, as they did not separately resolve a number of reflections for reciprocal lattice vectors {\bf G} in very different directions.  Later, Yamaguchi and Hashimoto \cite{yama72} performed magnetization measurements on a single crystal for ${\bf H} \parallel {\bf c}$ and ${\bf H}\perp{\bf c}$; they concluded that $c$ is the easy axis.  Our results indicate that neither of these conclusions is correct.

From Fig.~\ref{fg:200}, we see that there is significant FM intensity in both the (0 0 $-2$) and (2 0 0) reflections, from which we immediately conclude that there must be significant moment components perpendicular to both {\bf c} and {\bf a}.  The lack of temperature dependence of the (2 0 $-2$) intensity tells us that there cannot be a significant component along {\bf b}, consistent with the symmetry constraint.

\begin{figure}[t]
\includegraphics[width=0.9\columnwidth]{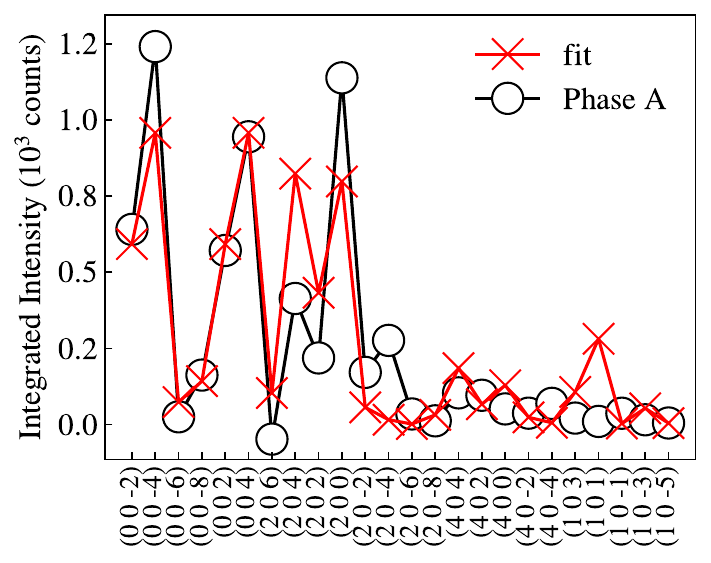}
\caption{\label{fg:FM_fit} Comparison of integrated intensity differences [$I$(4 K)\,$-$\,$I$(350 K)] (circles) and fitted results  (crosses) for phase A corresponding to the FM order at 4~K.}
\end{figure}

To quantitatively analyze the spin directions, we determined the integrated intensities for phase A at $T=4$~K and subtracted the values at 350~K to isolate the FM intensities.  We restricted the moment directions to be within the $a$-$c$ plane, but allowed the size and directions of the spins on the Cr$_{\rm h}$ and Cr$_{\rm i}$ sites to be independent.  Figure~\ref{fg:FM_fit} compares the experimental and fitted intensities.  The fit yields a moment of 5.6(3)~$\mu_{\rm B}$ on Cr$_{\rm h}$ oriented along $[0.72, 0, -0.69]$ (corresponding to an angle $\theta = 46^\circ$ from $-{\bf c}$ in the $a$-$c$ plane) and 3.7(5)~$\mu_{\rm B}$ on Cr$_{\rm i}$ oriented along $[0.90, 0, -0.45]$ ($\theta = 63^\circ$); the average moment per Cr is 5.0(4)~$\mu_{\rm B}$.  Both spin orientations are canted in the $a$-$c$ plane, but by different amounts.  %The average moment along the $c$ [$a$] axis per Cr atom is 3.2(2)~$\mu_{\rm B}$ [3.8(2)~$\mu_{\rm B}$].

%We should note, however, that volume of the neutron sample is not magnetically uniform; if we consider only crystals A and B, the ferromagnetic fraction is 76\%.  Taking that into account, we find a net $c$-axis moment of 2.4(2) $\mu_{\rm B}$, which is actually a bit below the measured saturation moment.
%Show evidence that ferromagnetic moment must have components along $a$ and $c$ and cannot be along $b$.

\begin{figure}[t]
 \centering
    \includegraphics[width=0.9\columnwidth]{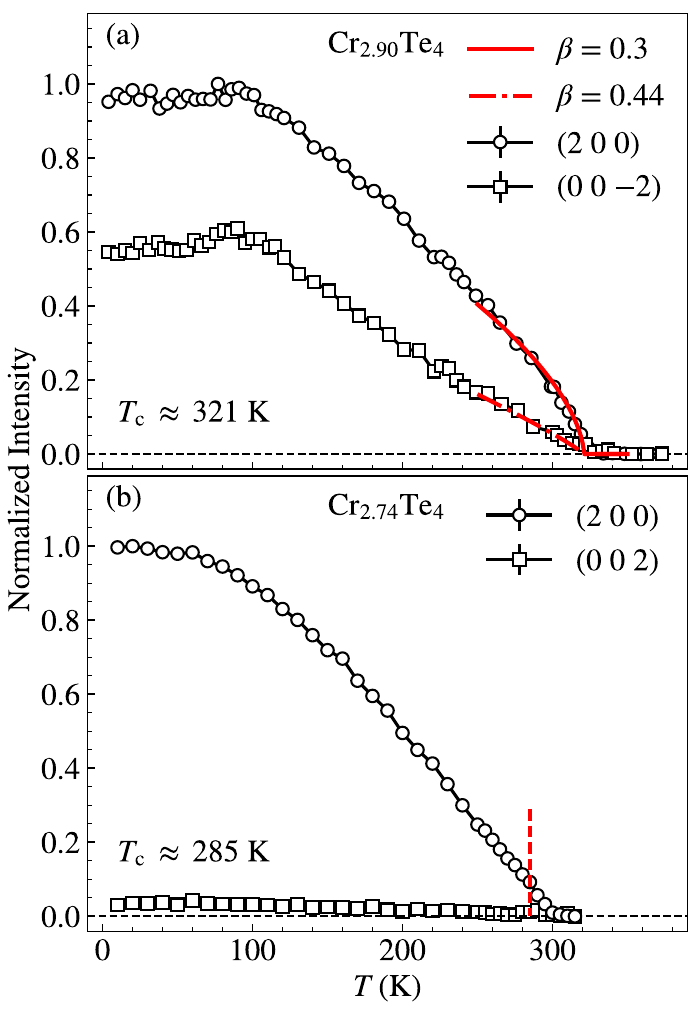}
    \caption{(a) Temperature dependence of the integrated rocking-curve intensities for the (2 0 0) [circles] and (0 0 $-2$) [squares] reflections for phase A of the $\delta = -0.10$ sample, where constant values observed at $T>321$~K have been subtracted and the intensities have been normalized to the (2 0 0) maximum.  The red lines are fits to $I\sim |T-T_{\rm C}|^{2\beta}$, where the resulting values of $\beta$ are listed in the legend. (b) Temperature dependence of integrated rocking-curve intensities for the (2 0 0) [circles] and (0 0 2) [squares] reflections of the $\delta = - 0.26$ sample, where constant values observed at $T>305$~K have been subtracted and the intensities have been normalized to the (2 0 0) maximum. %The integrated intensities for both the $(2,0,0)$ and $(0,0,2)$ reflections are normalized by $(2,0,0)$ reflection maxima.
    }
    \label{fg:FM_tdep} 
\end{figure}

The temperature dependencies of the FM intensities at (2 0 0) and (0 0 $-2$) are shown in Fig.~\ref{fg:FM_tdep}. The (2 0 0) appears to show critical behavior near $T_{\rm C} \approx 321$~K, the transition temperature identified from bulk magnetization measurements on a similar crystal \cite{gosw24a}, while the (0 0 $-2$) shows a distinct temperature dependence.  If we fit the intensities close to the transition with $I\sim |T-T_{\rm C}|^{2\beta}$, as indicated by the red lines, we obtain $\beta=0.30$ for (2 0 0) and $\beta=0.44$ for (0 0 $-2$).  It is interesting to note that these results bracket the value of $\beta=0.3827$ obtained in the magnetization study \cite{gosw24a}.  Thus, it appears that the character of the magnetic order is evolving near the transition.  In Sec.~\ref{sec:LS}, we will consider an interpretation in terms of the temperature dependence of the average spin orientation.  %This could involve a change in the average spin direction, a change in the relative moments on Cr$_{\rm h}$ and Cr$_{\rm i}$, or both.  We only followed these 2 peak intensities through the transition; more information would be required to resolve this question.

There is also a slight drop in the intensities of both peaks on cooling below $\sim90$~K, which corresponds with a similar change in the bulk magnetization measured both with field parallel and perpendicular to the $c$ axis \cite{gosw24a}.  One possible cause could be a temperature-dependent change in the relative canting of the moments on Cr$_{\rm h}$ and Cr$_{\rm i}$ sites. Another possible staggered spin canting would involve spin components of the Cr$_{\rm i}$ sites along the $b$ axis, with opposite orientations for the two sites per chemical unit cell; however, we can rule out that possibility, because it would yield a finite intensity at (0 0 1), whereas we could not detect any diffracted intensity at (0 0 1) or (0 0 $-1$).

\subsection{FM order in the $\delta=-0.26$ crystal}

The mosaic of the $\delta=-0.26$ crystal was broad and complex, so a complete crystallographic analysis did not seem practical.  Nevertheless, the character of the \gls*{FM} ordering is demonstrated by the temperature dependence of the (2 0 0) and (0 0 2) intensities, as shown in Fig.~\ref{fg:FM_tdep}(b).  As the \gls*{FM} intensity is strong for (2 0 0) and minimal for (0 0 2), it appears that the moments are aligned close to the $c$ axis.  There is no decrease in the FM signal at low temperature and we found no intensity at positions with half-integer $H$ or $L$ values, so there is no evidence for an AFM phase (though one cannot be absolutely ruled out).  This sample provides evidence that for larger densities of vacancies on the Cr$_{\rm i}$ site the FM order appears to be favored. 

\subsection{AFM phase from \gls*{SCND} for $\delta=-0.1$}
\label{sec:AFM}

\begin{figure}[t]
\includegraphics[width=0.9\columnwidth]{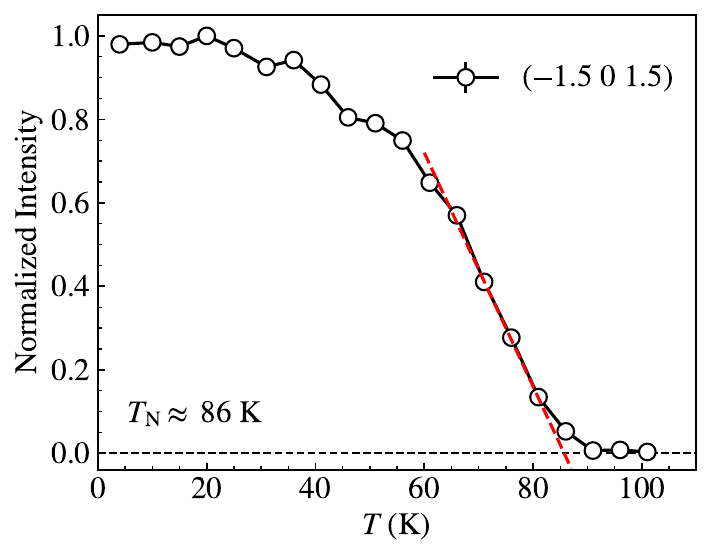}
\caption{\label{fg:AFM_T} Temperature dependence of the ($-1.5$ 0 1.5) peak of phase B of the $\delta=-0.10$ sample. The red line is a guide to eye and highlights the ordering temperature $T_\text{N} \approx$ 86 K. }
\end{figure}

\begin{figure}[b]
\includegraphics[width=0.99\columnwidth]{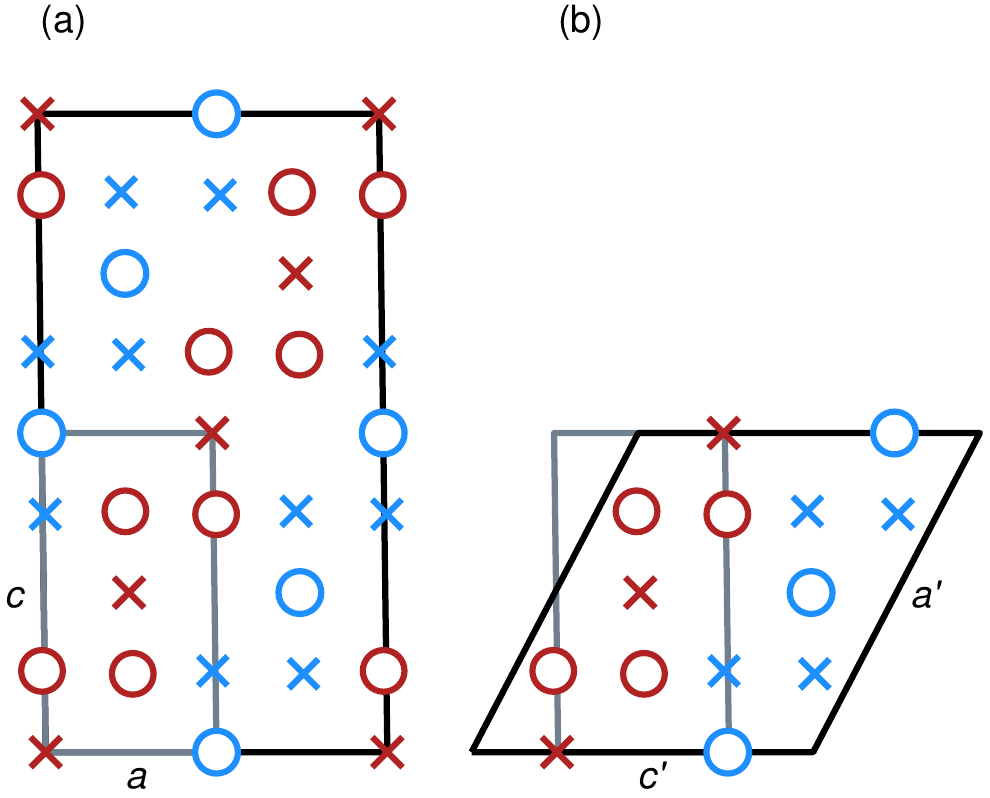}
\caption{\label{fg:AFM_model} Representations of the AFM order for two different choices of unit cell consistent with space group 12.  (a) AFM order within the $a$-$c$ plane, relative to the $I2/m$ unit cell, corresponding to model D of Bertaut {\it et al.} \cite{bert64}.  The $\circ$'s and {\sf x}'s represent heads and tails of arrows indicating spin direction along the $b$ axis. The red and blue highlight regions with opposite order of the AFM trimers.  With an ordering wave vector ${\bf k} = [\frac12, 0, -\frac12]$, the magnetic unit cell involves a quadrupling in size relative to the structural cell. (b) The proper minimal magnetic unit cell, based on magnetic space group $C_c2/m$ with ${\bf k} = [0, 0, \frac12]$. In both panels, the horizontal Cr$_{\rm h}$ layers are completely filled, while the Cr$_{\rm i}$ layers have vacancies.}
\end{figure}

The temperature dependence of the ($-1.5$ 0 1.5) reflection of phase B is shown in Fig.~\ref{fg:AFM_T}.  The extrapolation of the fitted line indicates $T_{\rm N} \approx 86$~K, which is compatible with the previous \gls*{NPD} studies \cite{bert64,andr70}.  

\begin{figure}[t]
\includegraphics[width=0.9\columnwidth]{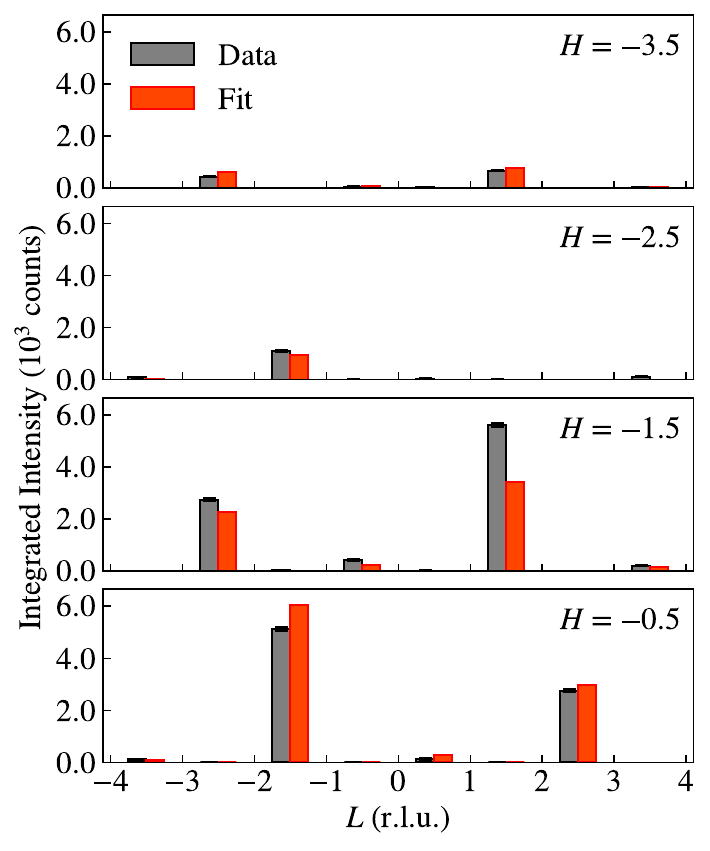}
\caption{\label{fg:AFM_fit} Comparison of measured integrated intensities of AFM reflections at 4~K and fitted results.  The error bars on the experimental data are quite small and hence are not shown.}
\end{figure}

We measured the integrated intensities of 17 AFM reflections for phase B at $T=4$~K.  To analyze the integrated intensities for these AFM peaks, we have used the model identified by Bertaut {\it et al.} \cite{bert64} and Andresen \cite{andr70} with the magnetic propagation wave vector of $[\frac12, 0, -\frac12]$, as  illustrated in Fig.~\ref{fg:AFM_model}(a). However, as discussed in Sec.~\ref{sec:ms}, the proper minimal magnetic unit cell, shown in Fig.~\ref{fg:AFM_model}(b), has just half the volume.  Figure~\ref{fg:AFM_fit} shows a comparison of measured and fitted integrated intensities.  Here we have assumed that the spins are oriented along the $b$ axis, which is consistent with the relevant magnetic space group; allowing canted spin components along {\bf a} or {\bf c} does not improve the fit (and would not be allowed by the chosen Irrep).  Note that our result for the spin direction is different from that of Andresen \cite{andr70}, who concluded that the spins are along ${\bf b}+{\bf c}$.
From the fit and the normalization to the nuclear intensities, we find that the the moments on both the Cr$_{\rm h}$ and Cr$_{\rm i}$ sites are large, corresponding to 3.8(4)~$\mu_{\rm B}$ and 3.6(4)~$\mu_{\rm B}$, respectively.

%Need a plot of AFM peaks along $(0.5,0,L)$ to show that these peaks work for B, but there is no $(0.5,0,1.5)_{\rm A}$ peak, as it should show up at $(0.5,0,-1.5)$.

%Note that AFM peaks require $H+L=$ even.

\begin{figure}[b]
    \centering
    \includegraphics[width=0.9\linewidth]{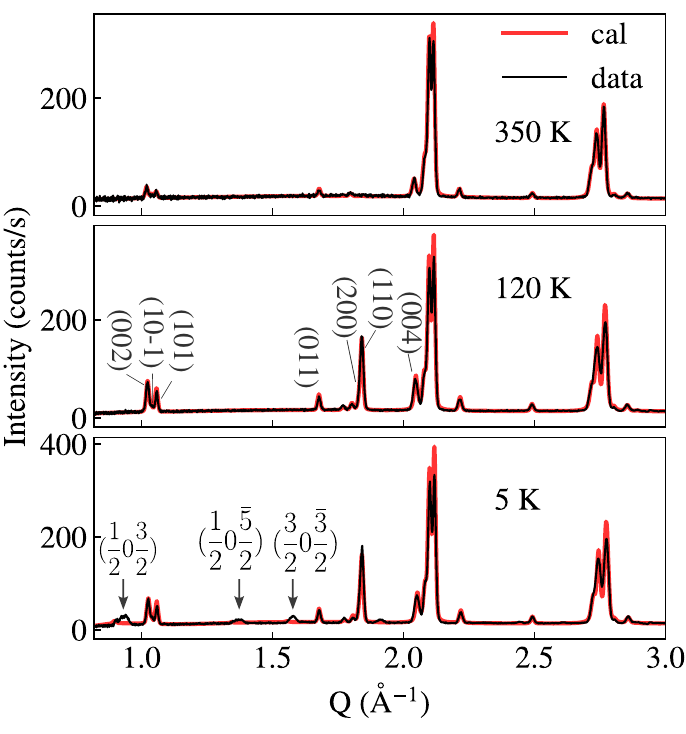}
    \caption{\gls*{NPD} data for the $\delta=-0.10$ sample at lower values of momentum transfer {\bf Q} to highlight the variations in intensities due to magnetic orders. Black arrows in the lower panel highlight peaks due to the AFM order from phase B. The $R_\text{wp}$ factors for 350 K, 120 K, and 5 K are 5.1$\%$, 6.4$\%$, and 6.9$\%$, respectively.}
    \label{fig:NPD_mag}
\end{figure}

\subsection{Magnetic moments from NPD}

To compare with the single-crystal results, we return to the \gls*{NPD} data and consider the refinement of the magnetic structures, along with the atomic order, in the low-temperature data. The magnetic scattering is strongest at low $Q$, so that is the range we compare for $T=5$, 120, and 350 K in Fig.~\ref{fig:NPD_mag}.  The fitting results for the atomic structure parameters are very similar to those at 350~K; for reference, the structural results at $T=5$~K are presented in the Appendix in Table~\ref{tb:struc5K}. The goodness of fit metrics for all temperatures satisfy  $R_{\rm wp}<7\%$.

%We used ISODISTORT \cite{Campbell_2006,Stokes_iso} to identify the appropriate magnetic space groups for the NPD analysis. The \gls*{FM} ($\mathbf{k} = (0\, 0\, 0)$ in $I2^\prime/m^\prime$ space group) and \gls*{AFM} ($\mathbf{k} = (0\, 0\, \frac{1}{2})$ in $C_c2/m$ space group) phases were introduced below their respective ordering temperatures. The magnetic space groups constrain the \gls*{FM} spins to the $ac-$plane and the \gls*{AFM} spins to the $b-$axis; these choices are made to be consistent with the single-crystal results.  
For the FM phase at 5~K, the magnetic moment of the Cr$_{\rm h}$ is described by the vector [1.7(1), 0, $-1.5(1)$], which corresponds to a magnitude of 2.3(1)~$\mu_{\rm B}$ pointing at an angle $\theta \approx 48 ^\circ$ from $-{\bf c}$; for Cr$_{\rm i}$ the results are 2.1(1)~$\mu_{\rm B}$ and $\theta \approx 55^\circ$.  These moment orientations are consistent with the single-crystal results when the uncertainties are taken into account. We have compiled the results for all three directions in Table~\ref{tb:npd}.

\begin{figure*}
\includegraphics[width=1.8\columnwidth]{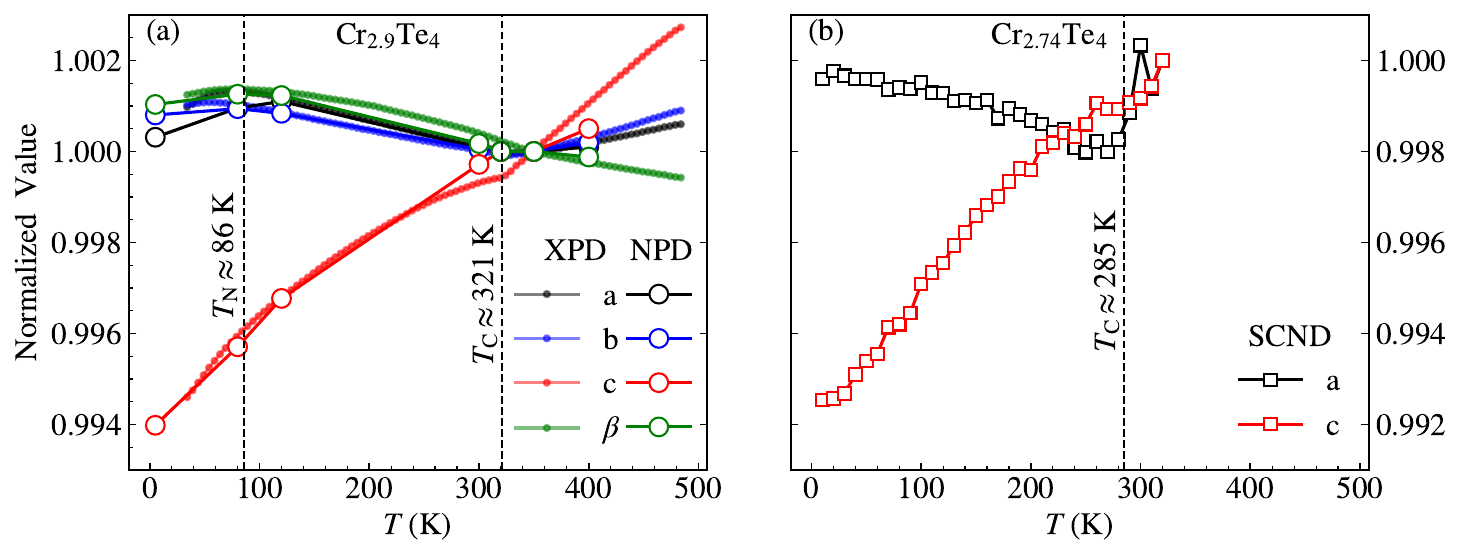}
%{figures/NPD_trend_norm_v2.pdf}

\caption{\label{fg:latt} Comparison of the thermal variations of the lattice parameters for the  $\delta = -0.10$ (a) and $\delta=-0.26$ (b) samples.  In panel (a) the filled and open circles represent Rietveld refinement results of \gls*{XPD} measured in temperature steps of 5 K and few representative \gls*{NPD}, respectively. The lattice parameter values have been normalized at 350~K.  The results in panel (b) are from $\theta$--$2\theta$ scans of \gls*{SCND} and normalized at 300~K.}
\end{figure*}

The moment values are normalized to the total volume of the sample.  If we assume that we have two phases present and the FM phase is 75.7\%\ of the volume, as for the single crystal sample, then the moments on the Cr$_{\rm h}$ and Cr$_{\rm i}$ sites at 5 K are 3.0(1) and 2.8(1)~$\mu_{\rm B}$, respectively, with an average of 2.9(1)~$\mu_{\rm B}$.  This average is only 58\%\ of that from the single-crystal analysis.

For the AFM phase, we found that there is a line-shape issue (that was not present in the single-crystal study).  For each expected AFM reflection, we observe a weak peak of the proper width at the appropriate position plus a broad peak shifted to larger $Q$.  While we do not have an interpretation of this behavior, we found that we could get a good fit to the integrated intensities if we take both Cr moments to be 0.9(1)~$\mu_{\rm B}$; normalizing to an AFM volume of 24\%, the average moment is 3.7(4), in good agreement with the single-crystal result.

\begin{table}[t]
    \caption{\label{tb:npd}  Results for the FM phase from the fits to the \gls*{NPD} data.  For the three measured temperatures, we list the magnetic moments and spin directions (angle from $-{\bf c}$) for the Cr$_{\rm h}$ and Cr$_{\rm i}$ and the average values weighted by the number of sites in the unit cell.}
\begin{ruledtabular}
    \begin{tabular}{dcccccc}
         T\ \  & $m_{\rm h}$ & $m_{\rm i}$ & $\langle m\rangle$ & $\theta_{\rm h}$ & $\theta_{\rm i}$ & $\langle \theta\rangle$ \\ 
        \ \ {\rm (K)} & ($\mu_{\rm B}$) & ($\mu_{\rm B}$) & ($\mu_{\rm B}$) & $(^\circ)$ & $(^\circ)$ & $(^\circ)$\rule[-5pt]{0pt}{12pt} \\
         \hline
         5 & 2.3(1) & 2.1(1) & 2.2(1) & 48(3) & 55(3) & 50(3)\rule[0pt]{0pt}{10pt}\\ 
         80 & 2.9(1) & 2.5(1) & 2.7(1) & 54(3) & 61(3) & 56(3) \\
         120 & 2.4(1) & 2.3(1) & 2.3(1) & 57(3) & 65(3) & 60(3) \\
\end{tabular}
\end{ruledtabular}
\end{table}

%These are the only possible space groups since the magnetic moment plane (\gls*{FM}) and direction (\gls*{AFM}) has been determined independently with \gls*{SCND}. 
%The magnetic moment magnitudes and orientation at the two Cr sites were the only magnetic parameters refined for each case; because we are limited to modeling a single structural phase, the magnetic moments of each phase are averaged over the entire sample volume. The \gls*{FM} moments at 5 K for Cr$_\text{h}$ [mx, my, mz] = [1.857(0.050) 0 $-$1.674(0.064)] $\mu_\text{B}$ and for Cr$_\text{i}$ = [1.971 (0.095) 0 $-$1.103(0.103)] $\mu_\text{B}$. For \gls*{AFM} phase the fit does not work well (R-factor is 20$\%$), producing Cr$_\text{h}$ (mx, my, mz) = [0 0.154(0.163) 0] $\mu_\text{B}$ and for Cr$_\text{i}$ = [0 -0.356(0.106) 0] $\mu_\text{B}$ small magnetic moment with large error bar. A few representative fitted \gls*{NPD} datasets are shown in Fig. and  the fitted lattice parameters from sequential fitting over all measured temperature are presented in Fig. \ref{fg:latt}(a). The goodness of global fit metrics for all temperatures were below R$_w$ $<$ 7$\%$.

\subsection{Lattice strain due to spontaneous magnetostriction}
\label{sec:LS}

\begin{figure}[b]
    \centering
    \includegraphics[width=0.9\linewidth]{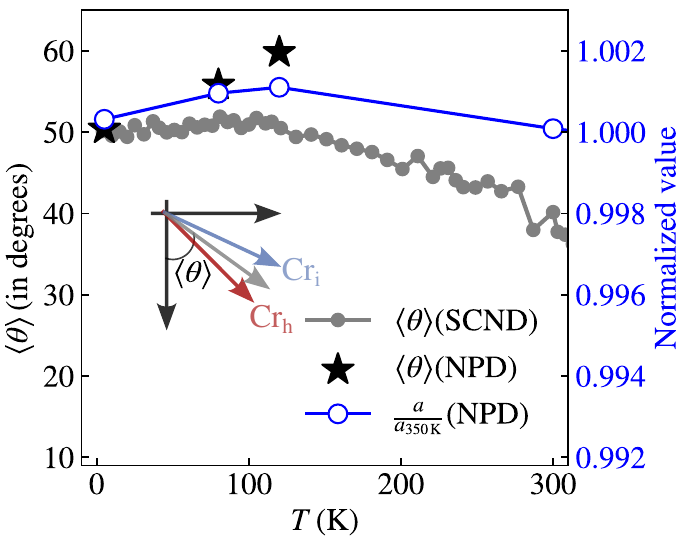}
    \caption{Temperature dependence of the average spin angle relative to $-{\bf c}$, $\langle\theta\rangle$, from \gls*{NPD} fits (stars) and estimated from \gls*{SCND} (2 0 0) and (0 0 $-2$) intensities (filled circles). Righthand axis shows normalized value of $a$ lattice parameter (open circles) with temperature, as in Fig.~\ref{fg:latt}(a).}
    \label{fig:spin_reorientation}
\end{figure}

Anisotropic lattice strain effects due to spontaneous magnetostriction have been reported for several compositions of Cr$_{1+x}$Te$_2$ in association with \gls*{FM} transitions \cite{li22}.  We have observed large effects in both of our samples, as shown in Fig.~\ref{fg:latt}.  We plot the temperature dependence of the lattice parameters normalized at a temperature above $T_{\rm C}$. On cooling through $T_{\rm C}$, for both samples, we see an abrupt change of slope in each lattice parameter relative to the paramagnetic phase.  This effect is so significant for $a$ and $b$ that we see negative thermal expansion in those directions.

\begin{figure}[b]
    \centering
    \includegraphics[width=0.9\linewidth]{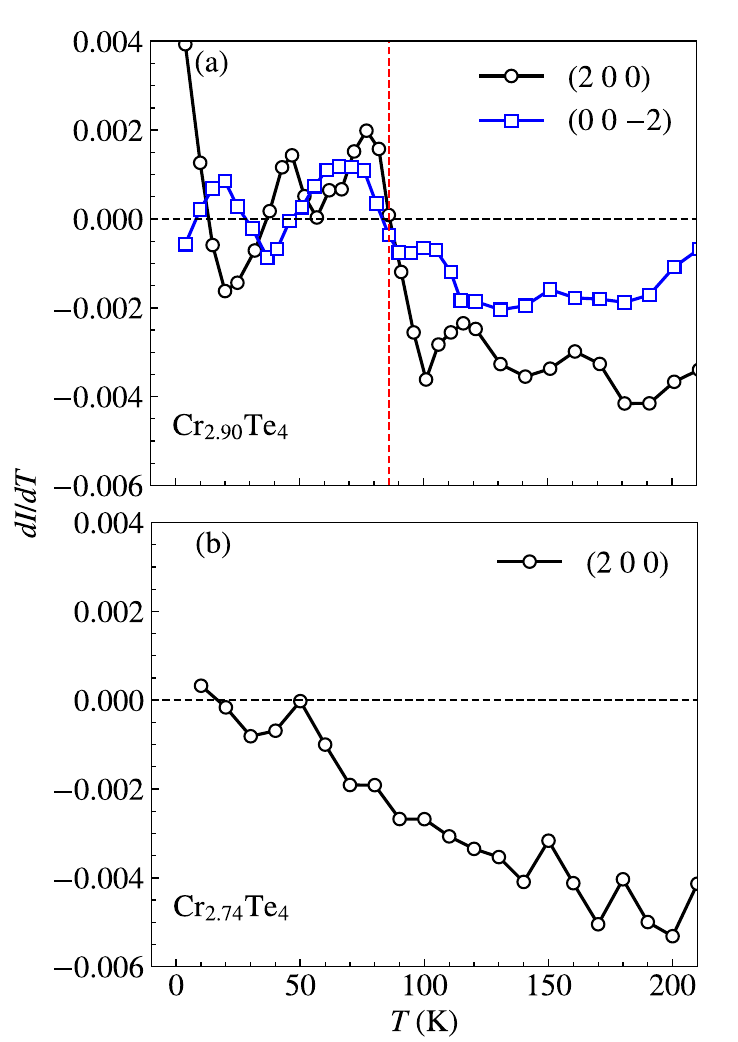}
    \caption{Temperature derivative of the integrated intensities (shown in Fig.~\ref{fg:FM_tdep}) for the (a) (2 0 0) and (0 0 $-2$) peaks of the $\delta=-0.1$ sample, and (b) the (2 0 0) peak of the $\delta=-0.26$ sample.  In both cases, a Savitzky-Golay filter was used to reduce the noise.  In (a), the vertical red dashed line is at $T=86$~K, which corresponds to $T_{\rm N}$ of phase B.}
    \label{fig:Ideriv}
\end{figure}

Intriguingly, the negative thermal expansion in $a$ and $b$ changes to positive thermal expansion below 100~K for the $\delta=-0.10$ sample but not for $\delta=-0.26$ [We did not measured $b$ for $\delta=-0.26$, as our scattering plane was ($H0L$)].  Furthermore, the former behavior is correlated with changes in the spin directions, as shown in Fig.~\ref{fig:spin_reorientation}.  From the NPD analysis, we found that the average spin direction  changes between 5 and 120~K, where the average is weighted by the number of atoms per unit cell and the ordered moments; we found that there was no significant change in the difference in spin directions for the two sites within the uncertainties.  To extend the comparison, we have made use of the single-crystal results in Fig.~\ref{fg:FM_tdep}(a); here we adjusted the normalization to be consistent with the average spin direction, then analyzed the $T$ dependence of the (2 0 0) and (0 0 $-2$) intensities assuming that the difference between them is due to the variation in average spin direction.  The results, average spin direction determined using \gls*{NPD} and \gls*{SCND}, show a clear correlation with the variation in the $a$ lattice parameter.  The spin directions are typically determined by exchange anisotropy from spin-orbit coupling, and it is understandable that this would be coupled to the lattice anisotropy.

An important question is why the thermal expansion of $a$ and $b$ changes sign below 100~K for the $\delta=-0.10$ sample.  This is close to $T_{\rm N} \approx 86$ K for the AFM order; however, that occurs in a separate phase B, distinct form the FM phase A.  Nevertheless, the calculations described in the next section indicate that the AFM phase prefers substantially smaller lattice parameters than those favored by the FM phase.  Also, we have seen that both phases appear to be present in a $\approx 100$ grain measured by TEM.  If domain widths of phases A and B are sufficiently small, then spontaneous magnetostriction in phase B will cause strain in phase A.  We propose that this is the explanation of the behavior in Fig.~\ref{fig:spin_reorientation}.

% If this idea of intertwined strains is correct, then we should expect the decrease in FM intensity seen at low temperature in Fig.~\ref{fg:FM_tdep} for phase A should onset at $T_{\rm N}\approx86$~K of phase B.  To test this, we plot in Fig.~\ref{fig:Ideriv} the temperature derivatives of the intensity data from Fig.~\ref{fg:FM_tdep}.  For the $\delta=-0.1$ sample, we see that $dI/dT$ passes through 0 right at $T_{\rm N}$.  In contrast, the results for the $\delta = -0.26$ show no clear sign of any strain-induced effect down to base temperature, consistent with the absence of an AFM phase in that sample.

If this idea of strain between the intertwined FM and AFM phases domain is correct, then we should expect the decrease in FM intensity at low temperature to onset at $T_{\rm N}$. This behavior is clearly visible in Fig.~\ref{fg:FM_tdep}(a), where lowering of FM phase A intensity occurs below $T_{\rm N}\approx86$~K of AFM phase B.  
To quantify this behavior, in Fig.~\ref{fig:Ideriv}, we plot  the temperature derivatives of the FM intensity data from Fig.~\ref{fg:FM_tdep}.  For the $\delta=-0.1$ sample, we see that $dI/dT$ passes through 0 right at $T_{\rm N}$.  In contrast, the results for the $\delta = -0.26$ show no clear sign of any strain-induced effect down to base temperature, consistent with the absence of an AFM phase in that sample.

\subsection{Analysis with \gls*{DFT}}
\label{sc:dft}

In an attempt to gain further insight, we performed calculations with \gls*{DFT}, in which all cell parameters were allowed to relax while maintaining the magnetic phase to be either FM or the AFM structure shown in Fig.~\ref{fg:AFM_model}(a).  In addition, we evaluated the impact of 1 or 2 vacancies on the Cr$_{\rm i}$ sublattice within the $2\times1\times2$ unit cell, corresponding to $\delta=-0.125$ and $-0.25$, respectively; in the case of 2 vacancies, they were placed in different interstitial layers.  The resulting unit cell parameters and energies per formula unit are presented in Table~\ref{tb:dft}.  Note that with one vacancy in the unit cell, the AFM phase becomes ferrimagnetic.

\begin{table}[b]
    \caption{\label{tb:dft} Lattice parameters and total energy for fully relaxed \gls*{FM} and \gls*{AFM} configurations for different Cr-vacancy concentrations in Cr$_{3+\delta}$Te$_4$ crystals. The energies for FM configuration are set to zero. }
\begin{ruledtabular}
    \begin{tabular}{l c c c c c c}%{dcddddd}
         $\delta$ & Phase & a & b &  c & $\beta$ & \multicolumn{1}{c}{$E$/{\rm f.u.}} \\ 
        & & & & & & \multicolumn{1}{c}{(meV)} \\
         \hline
         0 & FM  & 6.97 & 4.00 & 12.40 & 91.37 & 0 \\ %\hline 
         0 & AFM  & 6.82 & 3.98 & 12.36 & 91.02 & 161 \\ %\hline 
        $-0.125$ & FM  & 6.94 & 3.98 & 12.38 & 91.19 & 0 \\ %\hline 
         $-0.125$ & AFM  & 6.80 & 3.97 & 12.37 & 90.95 & 134 \\ %\hline 
         $-0.25$ & FM  & 6.90 & 3.97 & 12.35 & 90.98 & 0 \\ %\hline 
         $-0.25$ & AFM  & 6.75 & 3.97 & 12.28 & 90.84 & 108 \\ %\hline 
\end{tabular}
\end{ruledtabular}
\end{table}

For the fully-relaxed structures, the energy of the FM phase is always lower than that of the AFM, with the difference decreasing with vacancy density.  Clearly, DFT is not able to capture the stability of the AFM phase, even at $\delta=0$.  Given the large magnetic moments, perhaps this should not be surprising.  In any case, we note that the relaxed lattice parameters are quite different from the experimental values: both the FM and AFM unit cell volumes are larger than the experimental unit cell volume at 350 K (see caption of Table~\ref{tb:struc350} for lattice parameters).  Between the FM and AFM results for any value of $\delta$, the biggest difference occurs for $a$, which is 2\%\ larger for the FM phase. Comparing with experiment, we find $a_{\rm FM}>a_{\rm 350 K}>a_{\rm AFM}$.  This relationship is qualitatively consistent with the effects of spontaneous magnetostriction seen in experiment, as discussed in the last section.  When FM order develops in phase A on cooling, it prefers a larger value of $a$, as we observe in Fig.~\ref{fg:latt}.  When AFM order develops in phase B, it requires a smaller value of $a$, and so corresponding strain is generated.

As to what are the likely compositions of phases A and B in the $\delta=-0.10$ sample, 
the DFT results do not provide a simple answer, so we must rely on our empirical arguments.

\section{Summary and Discussion}
\label{sec:SD}

We have presented a variety of measurements with intertwined interpretations.  It may be helpful to start by summarizing our results before discussing their significance.  From neutron diffraction on a single-crystal sample of Cr$_{3-\delta}$Te$_4$ with $\delta=-0.10$ we observed two distinct phases, one of which develops FM order at 321 K and the other develops AFM order at 86 K.  %The two phases have their $c$ axes aligned but there $a$ axes in opposite directions. 
The two phases have their pseudo-hexagonal Cr layers in parallel orientations, but their ${\bf a}$ axes point in opposite directions; this means that the ${\bf c}$ axes of the two phases cant in opposite directions.
In contrast, the $\delta=-0.26$ crystal exhibits only FM order with $T_{\rm C}\approx 285$~K.

The magnetic structures observed are approximately consistent with those reported in early neutron powder diffraction studies \cite{bert64,andr70}.  For the FM phase, we are constrained to use the $I2'/m'$ magnetic space group, which allows the moments to point in the {\bf a}-{\bf c} plane.  From both the single crystal and the powder analyses, we find that the spins for both Cr$_{\rm h}$ and Cr$_{\rm i}$ sites are oriented near to the $[1, 0, -1]$ direction but somewhat closer to {\bf a}, with a difference of $\approx 10^\circ$ in orientation for the two sites; this differs from the orientation parallel to {\bf a} assumed by Andresen \cite{andr70}.  For the AFM phase, we have used the $C_c2/m$ magnetic space group, which is the only option consistent with the AFM propagation vector and which allows spins only along {\bf b}; this differs from Andresen's statement that the AFM moments point along ${\bf b}+{\bf c}$ \cite{andr70}.

The single crystal sample studied here is large, so there is the possibility that the two phases could be macroscopically separated.  Evidence for a finer scale segregation comes from the complementary measurements.  For example, the TEM measurement on an arbitrary grain with a thickness of $\approx100$~nm appears to provide structural evidence for the two phases.  Even more compelling are the correlations associated with the spontaneous magnetostriction.  As discussed in Sec.~\ref{sc:dft}, the changes in sign of the thermal variation of $a$ are consistent with predictions from DFT for the relative volumes of the FM and AFM phases.  The striking feature is that the FM moments and spin directions are correlated with the variations in $a$, including a reduction below $T_{\rm N}$, even though the AFM order is in a different part of the sample.  This interaction implies an effective epitaxial alignment of the two phases, with domain thicknesses that are sufficiently small that epitaxial strain effects are not mitigated by dislocations.

The AFM phase appears to have slightly larger values of $a$ and $\beta$ compared to the FM phase.  According to the composition-dependent study of Ipser {\it et al.} \cite{ipse83}, this implies that the Cr content of the AFM phase is slightly larger than that of the FM phase.  From the temperature vs.\ composition phase diagram \cite{ipse83}, the occurrence of spinodal decomposition as the crystal cools appears to be a possibility.  The internal strain of the crystal should limit the degree of compositional segregation and domain thickness.  Given that the crystal has an average of $\delta=-0.10$, we propose that the AFM phase may have $\delta \approx 0$, while the FM phase would correspond to 
$\delta\lesssim-0.1$.

%magnetostriction, correlation with magnetic changes, AFM impacts FM moments, TEM detection, fine-scale alignment of domains, spinodal decomposition, mention phase diagram \cite{ipse83}.  Mention other observations of spontaneous magnetostriction.

To evaluate the magnetic moments determined from the NPD analysis for $\delta=-0.10$, we have to estimate the relative volume fractions of the FM and AFM phases in the powder sample.  We have taken these to be the same as the relative volumes of phases A and B of the single crystal.  We should note that in doing this we have ignored the small phase C seen in Figs.~\ref{fg:rock} and \ref{fg:200}.  We believe that this represents a misaligned crystallite; it certainly contains a FM phase, but whether there is also an AFM phase is unclear.  In any case, it causes some uncertainty in the absolute volume fractions of the FM and AFM phases in the crystal.

With that understanding, the fitted magnitudes of the magnetic moments for the AFM phase showed reasonable agreement between the single-crystal and the powder results, with both yielding an average moment per Cr of 3.7~$\mu_{\rm B}$.  If we estimate the number of $3d$ electrons per Cr in a rough fashion, the valence $v$ of Cr ions in Cr$_{1+x}$Te$_2$ with Te$^{-2}$ should be $v=4/(1+x)$, yielding $v=2.67$ for $x=0.5$; the number of half-filled $3d$ levels should be equal to $6-v=3.33$.  Considering a spin of 1/2 per $3d$ electron and $g=2$, the expected moment would be 3.33~$\mu_{\rm B}$. Of course, there is good reason to expect spin-orbit coupling to be relevant \cite{purw23}, which can give $g>2$, and a value of $g\approx2.2$ would yield our experimental moment.

%; however, the results for the FM phase require some discussion.  The FM moments from the single-crystal analysis are significantly larger than powder results and larger than one would predict for spin-only moments (as we will discuss below).  Another measurement for comparison is bulk magnetization \cite{gosw24a}.

For the FM phase, it is relevant to compare the average moment with the saturation moment observed in bulk magnetization measurements on a single crystal \cite{gosw24a}. We expect that the AFM phase will make a negligible contribution. There is a complication, however.  At low fields, the magnetization shows a drop on cooling through $T_{\rm N}$. Correspondingly, we have observed a decrease in the average magnetic moment of the FM phase, as shown in Table~\ref{tb:npd}, which is correlated with the strain effect on the $a$ lattice parameter, as indicated in Fig.~\ref{fig:spin_reorientation}.  The challenge is that the drop in bulk magnetization largely disappears in high magnetic fields \cite{gosw24a,yama72}.  What is the significance of this change?

An answer is suggested by a recent study \cite{kubo23} of the impact of an applied field on the linear thermal expansion in a sintered polycrystalline sample of Cr$_3$Te$_4$. The applied field is generally observed to increase the length of the sample (in directions both parallel and perpendicular to the field), with the largest effects near $T_{\rm C}$ and below $T_{\rm N}$ for a field of 9~T \cite{kubo23}.  That result suggests that the field can compensate for the strain induced by the AFM phase. (Whether the field also suppresses the AFM order is an interesting question for a future experiment.)  Since the moment is correlated with the variations in $a$, it seems reasonable that the saturation moment should be larger than the zero-field NPD result at 5~K and possibly comparable to the moment we measured at 80~K.  Those average moments are 2.2 and 2.7~$\mu_{\rm B}$, respectively. The saturated moment from bulk magnetization at 9~T and 2~K is $\approx2.65~\mu_{\rm B}$, which is, indeed, above the neutron 5~K moment and close to the 80~K value.  (The average FM moment from the single-crystal analysis at 5~K, scaled by the FM volume fraction, corresponds to 3.8~$\mu_{\rm B}$, which is clearly too large; the source of the discrepancy is unclear.)

%To do this, we have to take account of the fact that phase B has only AFM, which will not make a significant contribution to the magnetization. 
%Assuming that our measured fractions of phases A and B apply to all crystals grown in the same batch, we find that the net moment per Cr atom is 2.4(2)~$\mu_{\rm B}$ [2.9(2)~$\mu_{\rm B}$] along the $c$ [$a$] axis. This compares with saturation moments at 2~K of 2.6 [2.5] $\mu_{\rm B}$ for $\mu_0H>4$~T.  Given the complications of the analysis, this is reasonable agreement.

%In the FM phase of crystal A, the fitted moment of 5.6~$\mu_{\rm B}$ for the Cr$_{\rm h}$ site seems unphysically large.   This is clearly much smaller than the fitted FM moment, although it is much closer to the fitted AFM moments.

Our observations on the correlation of the FM and AFM transitions with strain are consistent with reported measurements on a powder sample with $\delta=-0.06$ \cite{hata90}, where a pressure of 1~GPa caused $T_{\rm C}$ to decrease by 54 K and $T_{\rm N}$ to increase by 85 K.  The sensitivity to strain implies that thin films subject to epitaxial strain may have different ordering temperatures from bulk samples.  Indeed, a study of nonstoichiometric Cr$_3$Te$_4$ films grown on Al$_2$O$_3$ (001) substrates found that $T_{\rm C}$ grew from $\approx200$~K for a 3-layer film to $\approx300$~K for a film of $>20$ layers \cite{wang22}.

Studies of Cr$_{3+\delta}$Te$_4$ with angle-resolved photoemission \cite{fuji23,chal24} and band structure calculations \cite{dijk89,bose25} indicate the presence of Weyl points and possible contributions to the magnetism from Berry curvature.  Several studies of the anomalous Hall effect and changes below $T_{\rm N}$ have interpreted certain features in terms of topological effects due to chiral magnetism \cite{purw23,mats24,huan25}.  While we do see that the two Cr sites have slightly different spin directions in the FM phase, the spins are coplanar and, hence, not chiral. We propose that the effects below $T_{\rm N}$ should be attributed to FM/AFM multilayer effects rather than chiral magnetism within a single, uniform phase.  This raises an interesting question for future investigation: can multilayer films involving distinct nonchiral magnetic phases lead to an effective topological response in transport?

Our results may have relevance for other compositions of Cr$_{1+x}$Te$_2$.  For example, NPD measurements on a sample of monoclinic Cr$_5$Te$_8$ observed FM order with $T_{\rm C}=180$~K and AFM order with $T_{\rm N}=70$~K \cite{huan08}; however, no meaningful impact of the AFM ordering on the FM order parameter was apparent.  Do these orders occur within the same phase or in two slightly different compositions?  For similar compositions but with trigonal symmetry (presumably corresponding to a more disordered arrangement of the interstitial Cr), skyrmionic textures have been observed with Lorentz transmission electron microscopy for a temperature range between 100~K and 200~K \cite{saha22,prad24,rai25}.  These textures involve circular FM domains with diameters of $\approx300$~nm separated from the oppositely polarized background by N\'eel-type skyrmionic domain walls.  Could it be that disorder of interstitial Cr, with no symmetry constraints, allows more flexible and variable spin textures?  There have also been studies of crystals with similar composition in which the occurrence of an AFM phase in a range of 30--60~K above $T_{\rm C}$ was inferred from magnetization and magnetoresistance measurements \cite{zhan22,conn24}. It would certainly be of interest to test some of these behaviors with neutron scattering on single crystals.

\begin{table*}
\caption{Results of \gls*{NPD} analysis for the $\delta=-0.10$ sample at $T=5$~K using space group $I2/m$; the refined unit cell parameters are $a = 6.8563(2)$~\AA, $b = 3.9299(1)$~\AA, $c = 12.2352(4)$~\AA, and $\beta= 91.15(1)^\circ$.  \label{tb:struc5K}}
\begin{ruledtabular} 
\begin{tabular}{ccddddd} 
 Atom & site & x & y & z  & U & \multicolumn{1}{c}{\rm Occupancy} \\
  & & & & & \multicolumn{1}{c}{\rm (\AA$^2$)} & \\
\hline
Cr$_{\rm h}$ & 4$i$ & 0.9840(5) & 0 & 0.2565(3) & 0.0001(5) & 1 \\
Cr$_{\rm i}$ & 2$a$ & 0 & 0 & 0 & 0.0022(8) & 0.90 \\
%Cr$_{\rm i}$ & 2$c$ & 0 & 0 & 0.5 & 0.0118(4) & 0.107(5) \\
Te & 4$i$ & 0.6625(4) &  1 & 0.6321(2) & 0.0001 & 1 \\
Te & 4$i$ & 0.8351(3) & 0.5 & 0.3805(2) & 0.0001 & 1 \\
\end{tabular}
\end{ruledtabular}
\end{table*}

\begin{table*}
\caption{The AFM magnetic unit cell, in space group $C_c2/m$, of phase B in the $\delta=-0.10$ sample at $T=5$~K; the unit cell parameters are $a = 13.9878$~\AA, $b = 3.9299$~\AA, $c = 13.7228$~\AA, and $\beta= 118.29^\circ$.  
\label{tb:AFM_struc5K}} %The unit cell is predicted using ISODISTORT software suite using 350 K crystal structure extracted from \gls*{NPD}. 
\begin{ruledtabular} 
\begin{tabular}{ccddddd} 
 Atom & site & x & y & z  & U & \multicolumn{1}{c}{\rm Occupancy} \\
  & & & & & \multicolumn{1}{c}{\rm (\AA$^2$)} & \\
\hline
Cr$_{\rm h}$ & 4$i$ & 0.7446 & 0 & 0.8632 & 0.001 & 1 \\
Cr$_{\rm i}$ & 2$a$ & 0 & 0 & 0 & 0.001 & 1.0 \\
\end{tabular}
\end{ruledtabular}
\end{table*}

\section{Conclusion}

Our diffraction studies, especially on single-crystal samples, together with consistency between our results and those of past studies, lead us to the conclusion that crystals of Cr$_{3+\delta}$Te$_4$ with $\delta \approx 0$ are intrinsically segregated into thin lamellae of two distinct compositions, one phase with AFM order and $\delta\approx0$, and another with FM order and $\delta < - 0.1$.  The magnetic order in each phase is nonchiral; however, below $T_{\rm N}$, the net combination of magnetic orders across multiple lamellae is effectively chiral, which may explain topological effects detected in recent studies of the anomalous Hall effect \cite{purw23,mats24,huan25}.  Spontaneous magnetostriction of opposite signs for the AFM and FM phases results in interactions between the two phases via strain.  These results point to the need for further studies in order to properly understand the nature of the magnetic phases that occur in Cr$_{1+x}$Te$_2$ beyond $x\approx0.5$.

\section{Acknowledgments}
We thank B. Chakoumakos for a very helpful discussion.  Work at Brookhaven is supported by the Office of Basic Energy Sciences (BES), Materials Sciences and Engineering Division, U.S. Department of Energy (DOE) under Contract No.\ DE-SC0012704.
The work at Howard University is supported by the National Science Foundation Awards No.\ DMR-2018579 and No.\ DMR-2302436. This work used the synthesis facility of the Platform for the Accelerated Realization, Analysis, and Discovery of Interface Materials (PARADIM), which is supported by the National Science Foundation under Cooperative Agreement No.\ DMR-2039380.    This research also used resources at the High Flux Isotope Reactor and Spallation Neutron Source, DOE Office of Science User Facilities operated by ORNL. Beamtime was allocated to VERITAS on proposal numbers IPTS-30858 and IPTS-32807, and to POWGEN on proposal number IPTS-29079.

\appendix*
%\appendix
\section{Low temperature structural and magnetic refinement using NPD}

\begin{figure}[b]
\centering
\includegraphics[width=0.9\linewidth]{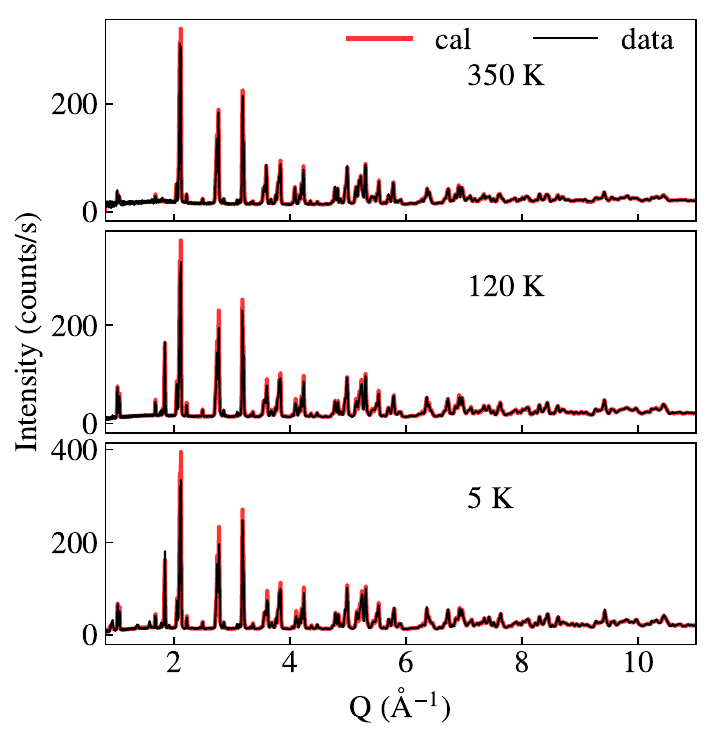}
\caption{Full range of the fitted \gls*{NPD} pattern at three different temperatures, subsections of which are shown in Fig.~\ref{fig:NPD_mag}.}
\label{fig:NPD_full_range}
\end{figure}

Here we present the results of the low-temperature refinements of crystal structure and ferromagnetic orders from the neutron powder diffraction data. The results of the structural refinement at $T=5$~K are listed in Table~\ref{tb:struc5K}.  The corresponding unit cell and parameters for the fit to the AFM order are presented in Table~\ref{tb:AFM_struc5K}.  

A comparison of NPD results and refinements for 3 temperatures and the full $Q$ range are compared in Fig.~\ref{fig:NPD_full_range}.
\vspace{6cm}
\bibliography{CrTe}

\begin{thebibliography}{36}%
\makeatletter
\providecommand \@ifxundefined [1]{%
 \@ifx{#1\undefined}
}%
\providecommand \@ifnum [1]{%
 \ifnum #1\expandafter \@firstoftwo
 \else \expandafter \@secondoftwo
 \fi
}%
\providecommand \@ifx [1]{%
 \ifx #1\expandafter \@firstoftwo
 \else \expandafter \@secondoftwo
 \fi
}%
\providecommand \natexlab [1]{#1}%
\providecommand \enquote  [1]{``#1''}%
\providecommand \bibnamefont  [1]{#1}%
\providecommand \bibfnamefont [1]{#1}%
\providecommand \citenamefont [1]{#1}%
\providecommand \href@noop [0]{\@secondoftwo}%
\providecommand \href [0]{\begingroup \@sanitize@url \@href}%
\providecommand \@href[1]{\@@startlink{#1}\@@href}%
\providecommand \@@href[1]{\endgroup#1\@@endlink}%
\providecommand \@sanitize@url [0]{\catcode `\\12\catcode `\$12\catcode `\&12\catcode `\#12\catcode `\^12\catcode `\_12\catcode `\%12\relax}%
\providecommand \@@startlink[1]{}%
\providecommand \@@endlink[0]{}%
\providecommand \url  [0]{\begingroup\@sanitize@url \@url }%
\providecommand \@url [1]{\endgroup\@href {#1}{\urlprefix }}%
\providecommand \urlprefix  [0]{URL }%
\providecommand \Eprint [0]{\href }%
\providecommand \doibase [0]{https://doi.org/}%
\providecommand \selectlanguage [0]{\@gobble}%
\providecommand \bibinfo  [0]{\@secondoftwo}%
\providecommand \bibfield  [0]{\@secondoftwo}%
\providecommand \translation [1]{[#1]}%
\providecommand \BibitemOpen [0]{}%
\providecommand \bibitemStop [0]{}%
\providecommand \bibitemNoStop [0]{.\EOS\space}%
\providecommand \EOS [0]{\spacefactor3000\relax}%
\providecommand \BibitemShut  [1]{\csname bibitem#1\endcsname}%
\let\auto@bib@innerbib\@empty
%</preamble>
\bibitem [{\citenamefont {Fujisawa}\ \emph {et~al.}(2020)\citenamefont {Fujisawa}, \citenamefont {Pardo-Almanza}, \citenamefont {Garland}, \citenamefont {Yamagami}, \citenamefont {Zhu}, \citenamefont {Chen}, \citenamefont {Araki}, \citenamefont {Takeda}, \citenamefont {Kobayashi}, \citenamefont {Takeda}, \citenamefont {Hsu}, \citenamefont {Chuang}, \citenamefont {Laskowski}, \citenamefont {Khoo}, \citenamefont {Soumyanarayanan},\ and\ \citenamefont {Okada}}]{fuji20}%
  \BibitemOpen
  \bibfield  {author} {\bibinfo {author} {\bibfnamefont {Y.}~\bibnamefont {Fujisawa}}, \bibinfo {author} {\bibfnamefont {M.}~\bibnamefont {Pardo-Almanza}}, \bibinfo {author} {\bibfnamefont {J.}~\bibnamefont {Garland}}, \bibinfo {author} {\bibfnamefont {K.}~\bibnamefont {Yamagami}}, \bibinfo {author} {\bibfnamefont {X.}~\bibnamefont {Zhu}}, \bibinfo {author} {\bibfnamefont {X.}~\bibnamefont {Chen}}, \bibinfo {author} {\bibfnamefont {K.}~\bibnamefont {Araki}}, \bibinfo {author} {\bibfnamefont {T.}~\bibnamefont {Takeda}}, \bibinfo {author} {\bibfnamefont {M.}~\bibnamefont {Kobayashi}}, \bibinfo {author} {\bibfnamefont {Y.}~\bibnamefont {Takeda}}, \bibinfo {author} {\bibfnamefont {C.~H.}\ \bibnamefont {Hsu}}, \bibinfo {author} {\bibfnamefont {F.~C.}\ \bibnamefont {Chuang}}, \bibinfo {author} {\bibfnamefont {R.}~\bibnamefont {Laskowski}}, \bibinfo {author} {\bibfnamefont {K.~H.}\ \bibnamefont {Khoo}}, \bibinfo {author} {\bibfnamefont {A.}~\bibnamefont {Soumyanarayanan}},\ and\ \bibinfo {author} {\bibfnamefont
  {Y.}~\bibnamefont {Okada}},\ }\bibfield  {title} {\bibinfo {title} {{Tailoring magnetism in self-intercalated Cr$_{1+\delta}$Te$_{2}$ epitaxial films}},\ }\href {https://doi.org/10.1103/PhysRevMaterials.4.114001} {\bibfield  {journal} {\bibinfo  {journal} {Phys. Rev. Mater.}\ }\textbf {\bibinfo {volume} {4}},\ \bibinfo {pages} {114001} (\bibinfo {year} {2020})}\BibitemShut {NoStop}%
\bibitem [{\citenamefont {Wang}\ \emph {et~al.}(2022)\citenamefont {Wang}, \citenamefont {Kajihara}, \citenamefont {Matsuoka}, \citenamefont {Saika}, \citenamefont {Yamagami}, \citenamefont {Takeda}, \citenamefont {Wadati}, \citenamefont {Ishizaka}, \citenamefont {Iwasa},\ and\ \citenamefont {Nakano}}]{wang22}%
  \BibitemOpen
  \bibfield  {author} {\bibinfo {author} {\bibfnamefont {Y.}~\bibnamefont {Wang}}, \bibinfo {author} {\bibfnamefont {S.}~\bibnamefont {Kajihara}}, \bibinfo {author} {\bibfnamefont {H.}~\bibnamefont {Matsuoka}}, \bibinfo {author} {\bibfnamefont {B.~K.}\ \bibnamefont {Saika}}, \bibinfo {author} {\bibfnamefont {K.}~\bibnamefont {Yamagami}}, \bibinfo {author} {\bibfnamefont {Y.}~\bibnamefont {Takeda}}, \bibinfo {author} {\bibfnamefont {H.}~\bibnamefont {Wadati}}, \bibinfo {author} {\bibfnamefont {K.}~\bibnamefont {Ishizaka}}, \bibinfo {author} {\bibfnamefont {Y.}~\bibnamefont {Iwasa}},\ and\ \bibinfo {author} {\bibfnamefont {M.}~\bibnamefont {Nakano}},\ }\bibfield  {title} {\bibinfo {title} {{Layer-Number-Independent Two-Dimensional Ferromagnetism in Cr$_3$Te$_4$}},\ }\href {https://doi.org/10.1021/acs.nanolett.2c03532} {\bibfield  {journal} {\bibinfo  {journal} {Nano Lett.}\ }\textbf {\bibinfo {volume} {22}},\ \bibinfo {pages} {9964} (\bibinfo {year} {2022})}\BibitemShut {NoStop}%
\bibitem [{\citenamefont {Matsuoka}\ \emph {et~al.}(2024)\citenamefont {Matsuoka}, \citenamefont {Kajihara}, \citenamefont {Nomoto}, \citenamefont {Wang}, \citenamefont {Hirayama}, \citenamefont {Arita}, \citenamefont {Iwasa},\ and\ \citenamefont {Nakano}}]{mats24}%
  \BibitemOpen
  \bibfield  {author} {\bibinfo {author} {\bibfnamefont {H.}~\bibnamefont {Matsuoka}}, \bibinfo {author} {\bibfnamefont {S.}~\bibnamefont {Kajihara}}, \bibinfo {author} {\bibfnamefont {T.}~\bibnamefont {Nomoto}}, \bibinfo {author} {\bibfnamefont {Y.}~\bibnamefont {Wang}}, \bibinfo {author} {\bibfnamefont {M.}~\bibnamefont {Hirayama}}, \bibinfo {author} {\bibfnamefont {R.}~\bibnamefont {Arita}}, \bibinfo {author} {\bibfnamefont {Y.}~\bibnamefont {Iwasa}},\ and\ \bibinfo {author} {\bibfnamefont {M.}~\bibnamefont {Nakano}},\ }\bibfield  {title} {\bibinfo {title} {{Band-driven switching of magnetism in a van der Waals magnetic semimetal}},\ }\href {https://doi.org/10.1126/sciadv.adk1415} {\bibfield  {journal} {\bibinfo  {journal} {Sci. Adv.}\ }\textbf {\bibinfo {volume} {10}},\ \bibinfo {pages} {eadk1415} (\bibinfo {year} {2024})}\BibitemShut {NoStop}%
\bibitem [{\citenamefont {Chaluvadi}\ \emph {et~al.}(2024)\citenamefont {Chaluvadi}, \citenamefont {Chalil}, \citenamefont {Jana}, \citenamefont {Dagur}, \citenamefont {Vinai}, \citenamefont {Motti}, \citenamefont {Fujii}, \citenamefont {Mezhoud}, \citenamefont {L{\"u}ders}, \citenamefont {Polewczyk}, \citenamefont {Vobornik}, \citenamefont {Rossi}, \citenamefont {Bigi}, \citenamefont {Hwang}, \citenamefont {Olsen}, \citenamefont {Orgiani},\ and\ \citenamefont {Mazzola}}]{chal24}%
  \BibitemOpen
  \bibfield  {author} {\bibinfo {author} {\bibfnamefont {S.~K.}\ \bibnamefont {Chaluvadi}}, \bibinfo {author} {\bibfnamefont {S.~P.}\ \bibnamefont {Chalil}}, \bibinfo {author} {\bibfnamefont {A.}~\bibnamefont {Jana}}, \bibinfo {author} {\bibfnamefont {D.}~\bibnamefont {Dagur}}, \bibinfo {author} {\bibfnamefont {G.}~\bibnamefont {Vinai}}, \bibinfo {author} {\bibfnamefont {F.}~\bibnamefont {Motti}}, \bibinfo {author} {\bibfnamefont {J.}~\bibnamefont {Fujii}}, \bibinfo {author} {\bibfnamefont {M.}~\bibnamefont {Mezhoud}}, \bibinfo {author} {\bibfnamefont {U.}~\bibnamefont {L{\"u}ders}}, \bibinfo {author} {\bibfnamefont {V.}~\bibnamefont {Polewczyk}}, \bibinfo {author} {\bibfnamefont {I.}~\bibnamefont {Vobornik}}, \bibinfo {author} {\bibfnamefont {G.}~\bibnamefont {Rossi}}, \bibinfo {author} {\bibfnamefont {C.}~\bibnamefont {Bigi}}, \bibinfo {author} {\bibfnamefont {Y.}~\bibnamefont {Hwang}}, \bibinfo {author} {\bibfnamefont {T.}~\bibnamefont {Olsen}}, \bibinfo {author} {\bibfnamefont {P.}~\bibnamefont
  {Orgiani}},\ and\ \bibinfo {author} {\bibfnamefont {F.}~\bibnamefont {Mazzola}},\ }\bibfield  {title} {\bibinfo {title} {{Uncovering the Lowest Thickness Limit for Room-Temperature Ferromagnetism of Cr$_{1.6}$Te$_2$}},\ }\href {https://doi.org/10.1021/acs.nanolett.4c01005} {\bibfield  {journal} {\bibinfo  {journal} {Nano Lett.}\ }\textbf {\bibinfo {volume} {24}},\ \bibinfo {pages} {7601} (\bibinfo {year} {2024})}\BibitemShut {NoStop}%
\bibitem [{\citenamefont {Guillet}\ \emph {et~al.}(2024)\citenamefont {Guillet}, \citenamefont {Boukari}, \citenamefont {Choueikani}, \citenamefont {Ohresser}, \citenamefont {Ouerghi}, \citenamefont {Mesple}, \citenamefont {Renard}, \citenamefont {Jacquot}, \citenamefont {Jalabert}, \citenamefont {Vergnaud}, \citenamefont {Bonell}, \citenamefont {Marty},\ and\ \citenamefont {Jamet}}]{guil24}%
  \BibitemOpen
  \bibfield  {author} {\bibinfo {author} {\bibfnamefont {Q.}~\bibnamefont {Guillet}}, \bibinfo {author} {\bibfnamefont {H.}~\bibnamefont {Boukari}}, \bibinfo {author} {\bibfnamefont {F.}~\bibnamefont {Choueikani}}, \bibinfo {author} {\bibfnamefont {P.}~\bibnamefont {Ohresser}}, \bibinfo {author} {\bibfnamefont {A.}~\bibnamefont {Ouerghi}}, \bibinfo {author} {\bibfnamefont {F.}~\bibnamefont {Mesple}}, \bibinfo {author} {\bibfnamefont {V.~T.}\ \bibnamefont {Renard}}, \bibinfo {author} {\bibfnamefont {J.-F.}\ \bibnamefont {Jacquot}}, \bibinfo {author} {\bibfnamefont {D.}~\bibnamefont {Jalabert}}, \bibinfo {author} {\bibfnamefont {C.}~\bibnamefont {Vergnaud}}, \bibinfo {author} {\bibfnamefont {F.}~\bibnamefont {Bonell}}, \bibinfo {author} {\bibfnamefont {A.}~\bibnamefont {Marty}},\ and\ \bibinfo {author} {\bibfnamefont {M.}~\bibnamefont {Jamet}},\ }\bibfield  {title} {\bibinfo {title} {{Magnetic evolution of Cr$_2$Te$_3$ epitaxially grown on graphene with post-growth annealing}},\ }\href
  {https://doi.org/10.1063/5.0200063} {\bibfield  {journal} {\bibinfo  {journal} {Appl. Phys. Lett.}\ }\textbf {\bibinfo {volume} {124}},\ \bibinfo {pages} {202402} (\bibinfo {year} {2024})}\BibitemShut {NoStop}%
\bibitem [{\citenamefont {Kushwaha}\ \emph {et~al.}(2025)\citenamefont {Kushwaha}, \citenamefont {Armitage}, \citenamefont {Edwards}, \citenamefont {Trzaska}, \citenamefont {Rigden}, \citenamefont {Bencok}, \citenamefont {Biswas}, \citenamefont {Lee}, \citenamefont {Sanders}, \citenamefont {van~der Laan}, \citenamefont {Wahl}, \citenamefont {King},\ and\ \citenamefont {Rajan}}]{kush25}%
  \BibitemOpen
  \bibfield  {author} {\bibinfo {author} {\bibfnamefont {N.}~\bibnamefont {Kushwaha}}, \bibinfo {author} {\bibfnamefont {O.}~\bibnamefont {Armitage}}, \bibinfo {author} {\bibfnamefont {B.}~\bibnamefont {Edwards}}, \bibinfo {author} {\bibfnamefont {L.}~\bibnamefont {Trzaska}}, \bibinfo {author} {\bibfnamefont {J.}~\bibnamefont {Rigden}}, \bibinfo {author} {\bibfnamefont {P.}~\bibnamefont {Bencok}}, \bibinfo {author} {\bibfnamefont {D.}~\bibnamefont {Biswas}}, \bibinfo {author} {\bibfnamefont {T.-L.}\ \bibnamefont {Lee}}, \bibinfo {author} {\bibfnamefont {C.}~\bibnamefont {Sanders}}, \bibinfo {author} {\bibfnamefont {G.}~\bibnamefont {van~der Laan}}, \bibinfo {author} {\bibfnamefont {P.}~\bibnamefont {Wahl}}, \bibinfo {author} {\bibfnamefont {P.~D.~C.}\ \bibnamefont {King}},\ and\ \bibinfo {author} {\bibfnamefont {A.}~\bibnamefont {Rajan}},\ }\bibfield  {title} {\bibinfo {title} {{From ferromagnetic semiconductor to antiferromagnetic metal in epitaxial Cr$_x$Te$_y$ monolayers}},\ }\href
  {https://doi.org/10.1038/s41535-025-00772-5} {\bibfield  {journal} {\bibinfo  {journal} {npj Quantum Mater.}\ }\textbf {\bibinfo {volume} {10}},\ \bibinfo {pages} {50} (\bibinfo {year} {2025})}\BibitemShut {NoStop}%
\bibitem [{\citenamefont {Ou}\ \emph {et~al.}(2025)\citenamefont {Ou}, \citenamefont {Mirzhalilov}, \citenamefont {Nemes}, \citenamefont {Martinez}, \citenamefont {Rocci}, \citenamefont {Duong}, \citenamefont {Akey}, \citenamefont {Foucher}, \citenamefont {Ge}, \citenamefont {Suri}, \citenamefont {Wang}, \citenamefont {Ambaye}, \citenamefont {Keum}, \citenamefont {Randeria}, \citenamefont {Trivedi}, \citenamefont {Burch}, \citenamefont {Bell}, \citenamefont {Ross}, \citenamefont {Wu}, \citenamefont {Heiman}, \citenamefont {Lauter}, \citenamefont {Moodera},\ and\ \citenamefont {Chi}}]{ou25}%
  \BibitemOpen
  \bibfield  {author} {\bibinfo {author} {\bibfnamefont {Y.}~\bibnamefont {Ou}}, \bibinfo {author} {\bibfnamefont {M.}~\bibnamefont {Mirzhalilov}}, \bibinfo {author} {\bibfnamefont {N.~M.}\ \bibnamefont {Nemes}}, \bibinfo {author} {\bibfnamefont {J.~L.}\ \bibnamefont {Martinez}}, \bibinfo {author} {\bibfnamefont {M.}~\bibnamefont {Rocci}}, \bibinfo {author} {\bibfnamefont {A.}~\bibnamefont {Duong}}, \bibinfo {author} {\bibfnamefont {A.}~\bibnamefont {Akey}}, \bibinfo {author} {\bibfnamefont {A.~C.}\ \bibnamefont {Foucher}}, \bibinfo {author} {\bibfnamefont {W.}~\bibnamefont {Ge}}, \bibinfo {author} {\bibfnamefont {D.}~\bibnamefont {Suri}}, \bibinfo {author} {\bibfnamefont {Y.}~\bibnamefont {Wang}}, \bibinfo {author} {\bibfnamefont {H.}~\bibnamefont {Ambaye}}, \bibinfo {author} {\bibfnamefont {J.}~\bibnamefont {Keum}}, \bibinfo {author} {\bibfnamefont {M.}~\bibnamefont {Randeria}}, \bibinfo {author} {\bibfnamefont {N.}~\bibnamefont {Trivedi}}, \bibinfo {author} {\bibfnamefont {K.~S.}\ \bibnamefont {Burch}},
  \bibinfo {author} {\bibfnamefont {D.~C.}\ \bibnamefont {Bell}}, \bibinfo {author} {\bibfnamefont {F.~M.}\ \bibnamefont {Ross}}, \bibinfo {author} {\bibfnamefont {W.}~\bibnamefont {Wu}}, \bibinfo {author} {\bibfnamefont {D.}~\bibnamefont {Heiman}}, \bibinfo {author} {\bibfnamefont {V.}~\bibnamefont {Lauter}}, \bibinfo {author} {\bibfnamefont {J.~S.}\ \bibnamefont {Moodera}},\ and\ \bibinfo {author} {\bibfnamefont {H.}~\bibnamefont {Chi}},\ }\bibfield  {title} {\bibinfo {title} {{Enhanced ferromagnetism in monolayer Cr$_2$Te$_3$ via topological insulator coupling}},\ }\href {https://doi.org/10.1088/1361-6633/add9c5} {\bibfield  {journal} {\bibinfo  {journal} {Rep. Prog. Phys.}\ }\textbf {\bibinfo {volume} {88}},\ \bibinfo {pages} {060501} (\bibinfo {year} {2025})}\BibitemShut {NoStop}%
\bibitem [{\citenamefont {He}\ \emph {et~al.}(2025)\citenamefont {He}, \citenamefont {Bian}, \citenamefont {Seddon}, \citenamefont {Jagadish}, \citenamefont {Mucchietto}, \citenamefont {Ren}, \citenamefont {Kirstein}, \citenamefont {Asadi}, \citenamefont {Bai}, \citenamefont {Yao}, \citenamefont {Pan}, \citenamefont {Yu}, \citenamefont {Milde}, \citenamefont {Huai}, \citenamefont {Hui}, \citenamefont {Zang}, \citenamefont {Sabirianov}, \citenamefont {Cheng}, \citenamefont {Miao}, \citenamefont {Xing}, \citenamefont {Shao}, \citenamefont {Crooker}, \citenamefont {Eng}, \citenamefont {Hou}, \citenamefont {Bird},\ and\ \citenamefont {Zeng}}]{he25}%
  \BibitemOpen
  \bibfield  {author} {\bibinfo {author} {\bibfnamefont {K.}~\bibnamefont {He}}, \bibinfo {author} {\bibfnamefont {M.}~\bibnamefont {Bian}}, \bibinfo {author} {\bibfnamefont {S.~D.}\ \bibnamefont {Seddon}}, \bibinfo {author} {\bibfnamefont {K.}~\bibnamefont {Jagadish}}, \bibinfo {author} {\bibfnamefont {A.}~\bibnamefont {Mucchietto}}, \bibinfo {author} {\bibfnamefont {H.}~\bibnamefont {Ren}}, \bibinfo {author} {\bibfnamefont {E.}~\bibnamefont {Kirstein}}, \bibinfo {author} {\bibfnamefont {R.}~\bibnamefont {Asadi}}, \bibinfo {author} {\bibfnamefont {J.}~\bibnamefont {Bai}}, \bibinfo {author} {\bibfnamefont {C.}~\bibnamefont {Yao}}, \bibinfo {author} {\bibfnamefont {S.}~\bibnamefont {Pan}}, \bibinfo {author} {\bibfnamefont {J.-X.}\ \bibnamefont {Yu}}, \bibinfo {author} {\bibfnamefont {P.}~\bibnamefont {Milde}}, \bibinfo {author} {\bibfnamefont {C.}~\bibnamefont {Huai}}, \bibinfo {author} {\bibfnamefont {H.}~\bibnamefont {Hui}}, \bibinfo {author} {\bibfnamefont {J.}~\bibnamefont {Zang}}, \bibinfo {author}
  {\bibfnamefont {R.}~\bibnamefont {Sabirianov}}, \bibinfo {author} {\bibfnamefont {X.~M.}\ \bibnamefont {Cheng}}, \bibinfo {author} {\bibfnamefont {G.}~\bibnamefont {Miao}}, \bibinfo {author} {\bibfnamefont {H.}~\bibnamefont {Xing}}, \bibinfo {author} {\bibfnamefont {Y.-T.}\ \bibnamefont {Shao}}, \bibinfo {author} {\bibfnamefont {S.~A.}\ \bibnamefont {Crooker}}, \bibinfo {author} {\bibfnamefont {L.}~\bibnamefont {Eng}}, \bibinfo {author} {\bibfnamefont {Y.}~\bibnamefont {Hou}}, \bibinfo {author} {\bibfnamefont {J.~P.}\ \bibnamefont {Bird}},\ and\ \bibinfo {author} {\bibfnamefont {H.}~\bibnamefont {Zeng}},\ }\bibfield  {title} {\bibinfo {title} {{Unconventional Anomalous Hall Effect Driven by Self-Intercalation in Covalent 2D Magnet Cr$_2$Te$_3$}},\ }\href {https://doi.org/https://doi.org/10.1002/advs.202407625} {\bibfield  {journal} {\bibinfo  {journal} {Adv. Sci.}\ }\textbf {\bibinfo {volume} {12}},\ \bibinfo {pages} {2407625} (\bibinfo {year} {2025})}\BibitemShut {NoStop}%
\bibitem [{\citenamefont {Ipser}\ \emph {et~al.}(1983)\citenamefont {Ipser}, \citenamefont {Komarek},\ and\ \citenamefont {Klepp}}]{ipse83}%
  \BibitemOpen
  \bibfield  {author} {\bibinfo {author} {\bibfnamefont {H.}~\bibnamefont {Ipser}}, \bibinfo {author} {\bibfnamefont {K.~L.}\ \bibnamefont {Komarek}},\ and\ \bibinfo {author} {\bibfnamefont {K.~O.}\ \bibnamefont {Klepp}},\ }\bibfield  {title} {\bibinfo {title} {{Transition metal-chalcogen systems viii: The Cr--Te phase diagram}},\ }\href {https://doi.org/https://doi.org/10.1016/0022-5088(83)90493-9} {\bibfield  {journal} {\bibinfo  {journal} {J. Less Common Met.}\ }\textbf {\bibinfo {volume} {92}},\ \bibinfo {pages} {265} (\bibinfo {year} {1983})}\BibitemShut {NoStop}%
\bibitem [{\citenamefont {{Bertaut, E.F.}}\ \emph {et~al.}(1964)\citenamefont {{Bertaut, E.F.}}, \citenamefont {{Roult, G.}}, \citenamefont {{Aleonard, R.}}, \citenamefont {{Pauthenet, R.}}, \citenamefont {{Chevreton, M.}},\ and\ \citenamefont {{Jansen, R.}}}]{bert64}%
  \BibitemOpen
  \bibfield  {author} {\bibinfo {author} {\bibnamefont {{Bertaut, E.F.}}}, \bibinfo {author} {\bibnamefont {{Roult, G.}}}, \bibinfo {author} {\bibnamefont {{Aleonard, R.}}}, \bibinfo {author} {\bibnamefont {{Pauthenet, R.}}}, \bibinfo {author} {\bibnamefont {{Chevreton, M.}}},\ and\ \bibinfo {author} {\bibnamefont {{Jansen, R.}}},\ }\bibfield  {title} {\bibinfo {title} {{Structures magn\'etiques de Cr$_3$X$_4$ (X = S, Se, Te)}},\ }\href {https://doi.org/10.1051/jphys:01964002505058200} {\bibfield  {journal} {\bibinfo  {journal} {J. Phys. France}\ }\textbf {\bibinfo {volume} {25}},\ \bibinfo {pages} {582} (\bibinfo {year} {1964})}\BibitemShut {NoStop}%
\bibitem [{\citenamefont {Andresen}(1970)}]{andr70}%
  \BibitemOpen
  \bibfield  {author} {\bibinfo {author} {\bibfnamefont {A.~F.}\ \bibnamefont {Andresen}},\ }\bibfield  {title} {\bibinfo {title} {{The Magnetic Structure of Cr$_2$Te$_3$, Cr$_3$Te$_4$, and Cr$_5$Te$_6$}},\ }\href {https://doi.org/10.3891/acta.chem.scand.24-3495} {\bibfield  {journal} {\bibinfo  {journal} {Acta Chem. Scand.}\ }\textbf {\bibinfo {volume} {24}},\ \bibinfo {pages} {3495} (\bibinfo {year} {1970})}\BibitemShut {NoStop}%
\bibitem [{\citenamefont {Yamaguchi}\ and\ \citenamefont {Hashimoto}(1972)}]{yama72}%
  \BibitemOpen
  \bibfield  {author} {\bibinfo {author} {\bibfnamefont {M.}~\bibnamefont {Yamaguchi}}\ and\ \bibinfo {author} {\bibfnamefont {T.}~\bibnamefont {Hashimoto}},\ }\bibfield  {title} {\bibinfo {title} {{Magnetic Properties of Cr$_3$Te$_4$ in Ferromagnetic Region}},\ }\href {https://doi.org/10.1143/JPSJ.32.635} {\bibfield  {journal} {\bibinfo  {journal} {J. Phys. Soc. Jpn.}\ }\textbf {\bibinfo {volume} {32}},\ \bibinfo {pages} {635} (\bibinfo {year} {1972})}\BibitemShut {NoStop}%
\bibitem [{\citenamefont {Goswami}\ \emph {et~al.}(2024)\citenamefont {Goswami}, \citenamefont {Ng}, \citenamefont {Yakubu}, \citenamefont {Abeykoon},\ and\ \citenamefont {Guchhait}}]{gosw24a}%
  \BibitemOpen
  \bibfield  {author} {\bibinfo {author} {\bibfnamefont {A.}~\bibnamefont {Goswami}}, \bibinfo {author} {\bibfnamefont {N.}~\bibnamefont {Ng}}, \bibinfo {author} {\bibfnamefont {E.}~\bibnamefont {Yakubu}}, \bibinfo {author} {\bibfnamefont {A.~M.}\ \bibnamefont {Abeykoon}},\ and\ \bibinfo {author} {\bibfnamefont {S.}~\bibnamefont {Guchhait}},\ }\bibfield  {title} {\bibinfo {title} {{Critical behavior in monoclinic ${\mathrm{Cr}}_{3}{\mathrm{Te}}_{4}$}},\ }\href {https://doi.org/10.1103/PhysRevB.109.054413} {\bibfield  {journal} {\bibinfo  {journal} {Phys. Rev. B}\ }\textbf {\bibinfo {volume} {109}},\ \bibinfo {pages} {054413} (\bibinfo {year} {2024})}\BibitemShut {NoStop}%
\bibitem [{\citenamefont {R\"oseler}\ \emph {et~al.}(2025)\citenamefont {R\"oseler}, \citenamefont {Witteveen}, \citenamefont {Besnard}, \citenamefont {Pomjakushin}, \citenamefont {Jeschke},\ and\ \citenamefont {von Rohr}}]{rose25}%
  \BibitemOpen
  \bibfield  {author} {\bibinfo {author} {\bibfnamefont {K.~D.}\ \bibnamefont {R\"oseler}}, \bibinfo {author} {\bibfnamefont {C.}~\bibnamefont {Witteveen}}, \bibinfo {author} {\bibfnamefont {C.}~\bibnamefont {Besnard}}, \bibinfo {author} {\bibfnamefont {V.}~\bibnamefont {Pomjakushin}}, \bibinfo {author} {\bibfnamefont {H.~O.}\ \bibnamefont {Jeschke}},\ and\ \bibinfo {author} {\bibfnamefont {F.~O.}\ \bibnamefont {von Rohr}},\ }\bibfield  {title} {\bibinfo {title} {{Efficient soft-chemical synthesis of large van-der-Waals crystals of the room-temperature ferromagnet 1T-CrTe$_2$}},\ }\href {https://doi.org/10.1039/D4TA05649C} {\bibfield  {journal} {\bibinfo  {journal} {J. Mater. Chem. A}\ }\textbf {\bibinfo {volume} {13}},\ \bibinfo {pages} {15798} (\bibinfo {year} {2025})}\BibitemShut {NoStop}%
\bibitem [{\citenamefont {Liu}\ \emph {et~al.}(2019)\citenamefont {Liu}, \citenamefont {Abeykoon}, \citenamefont {Stavitski}, \citenamefont {Attenkofer},\ and\ \citenamefont {Petrovic}}]{liu19}%
  \BibitemOpen
  \bibfield  {author} {\bibinfo {author} {\bibfnamefont {Y.}~\bibnamefont {Liu}}, \bibinfo {author} {\bibfnamefont {M.}~\bibnamefont {Abeykoon}}, \bibinfo {author} {\bibfnamefont {E.}~\bibnamefont {Stavitski}}, \bibinfo {author} {\bibfnamefont {K.}~\bibnamefont {Attenkofer}},\ and\ \bibinfo {author} {\bibfnamefont {C.}~\bibnamefont {Petrovic}},\ }\bibfield  {title} {\bibinfo {title} {{Magnetic anisotropy and entropy change in trigonal ${\mathrm{Cr}}_{5}{\mathrm{Te}}_{8}$}},\ }\href {https://doi.org/10.1103/PhysRevB.100.245114} {\bibfield  {journal} {\bibinfo  {journal} {Phys. Rev. B}\ }\textbf {\bibinfo {volume} {100}},\ \bibinfo {pages} {245114} (\bibinfo {year} {2019})}\BibitemShut {NoStop}%
\bibitem [{\citenamefont {Lotgering}\ and\ \citenamefont {Gorter}(1957)}]{lotg57}%
  \BibitemOpen
  \bibfield  {author} {\bibinfo {author} {\bibfnamefont {F.}~\bibnamefont {Lotgering}}\ and\ \bibinfo {author} {\bibfnamefont {E.}~\bibnamefont {Gorter}},\ }\bibfield  {title} {\bibinfo {title} {{Solid solutions between ferromagnetic and antiferromagnetic compounds with NiAs structure}},\ }\href {https://doi.org/https://doi.org/10.1016/0022-3697(57)90028-8} {\bibfield  {journal} {\bibinfo  {journal} {J. Phys. Chem. Solids}\ }\textbf {\bibinfo {volume} {3}},\ \bibinfo {pages} {238} (\bibinfo {year} {1957})}\BibitemShut {NoStop}%
\bibitem [{\citenamefont {Huang}\ \emph {et~al.}(2008)\citenamefont {Huang}, \citenamefont {Kockelmann}, \citenamefont {Telling},\ and\ \citenamefont {Bensch}}]{huan08}%
  \BibitemOpen
  \bibfield  {author} {\bibinfo {author} {\bibfnamefont {Z.-L.}\ \bibnamefont {Huang}}, \bibinfo {author} {\bibfnamefont {W.}~\bibnamefont {Kockelmann}}, \bibinfo {author} {\bibfnamefont {M.}~\bibnamefont {Telling}},\ and\ \bibinfo {author} {\bibfnamefont {W.}~\bibnamefont {Bensch}},\ }\bibfield  {title} {\bibinfo {title} {{A neutron diffraction study of structural and magnetic properties of monoclinic Cr$_5$Te$_8$}},\ }\href {https://doi.org/https://doi.org/10.1016/j.solidstatesciences.2007.11.013} {\bibfield  {journal} {\bibinfo  {journal} {Solid State Sci.}\ }\textbf {\bibinfo {volume} {10}},\ \bibinfo {pages} {1099} (\bibinfo {year} {2008})}\BibitemShut {NoStop}%
\bibitem [{\citenamefont {Ohsawa}\ \emph {et~al.}(1972)\citenamefont {Ohsawa}, \citenamefont {Yamaguchi}, \citenamefont {Kazama}, \citenamefont {Yamauchi},\ and\ \citenamefont {Watanabe}}]{ohsa72}%
  \BibitemOpen
  \bibfield  {author} {\bibinfo {author} {\bibfnamefont {A.}~\bibnamefont {Ohsawa}}, \bibinfo {author} {\bibfnamefont {Y.}~\bibnamefont {Yamaguchi}}, \bibinfo {author} {\bibfnamefont {N.}~\bibnamefont {Kazama}}, \bibinfo {author} {\bibfnamefont {H.}~\bibnamefont {Yamauchi}},\ and\ \bibinfo {author} {\bibfnamefont {H.}~\bibnamefont {Watanabe}},\ }\bibfield  {title} {\bibinfo {title} {{Magnetic Anisotropy of Cr$_{1-x}$Te with $x=0.077$}},\ }\href {https://doi.org/10.1143/JPSJ.33.1303} {\bibfield  {journal} {\bibinfo  {journal} {J. Phys. Soc. Jpn.}\ }\textbf {\bibinfo {volume} {33}},\ \bibinfo {pages} {1303} (\bibinfo {year} {1972})}\BibitemShut {NoStop}%
\bibitem [{\citenamefont {Hamasaki}\ \emph {et~al.}(1975)\citenamefont {Hamasaki}, \citenamefont {Hashimoto}, \citenamefont {Yamaguchi},\ and\ \citenamefont {Watanabe}}]{hama75}%
  \BibitemOpen
  \bibfield  {author} {\bibinfo {author} {\bibfnamefont {T.}~\bibnamefont {Hamasaki}}, \bibinfo {author} {\bibfnamefont {T.}~\bibnamefont {Hashimoto}}, \bibinfo {author} {\bibfnamefont {Y.}~\bibnamefont {Yamaguchi}},\ and\ \bibinfo {author} {\bibfnamefont {H.}~\bibnamefont {Watanabe}},\ }\bibfield  {title} {\bibinfo {title} {{Neutron diffraction study of Cr$_2$Te$_3$ single crystal}},\ }\href {https://doi.org/https://doi.org/10.1016/0038-1098(75)90888-1} {\bibfield  {journal} {\bibinfo  {journal} {Solid State Commun.}\ }\textbf {\bibinfo {volume} {16}},\ \bibinfo {pages} {895} (\bibinfo {year} {1975})}\BibitemShut {NoStop}%
\bibitem [{\citenamefont {Cox}\ \emph {et~al.}(1965)\citenamefont {Cox}, \citenamefont {Shirane},\ and\ \citenamefont {Takei}}]{cox65}%
  \BibitemOpen
  \bibfield  {author} {\bibinfo {author} {\bibfnamefont {D.~E.}\ \bibnamefont {Cox}}, \bibinfo {author} {\bibfnamefont {G.}~\bibnamefont {Shirane}},\ and\ \bibinfo {author} {\bibfnamefont {W.~J.}\ \bibnamefont {Takei}},\ }\bibfield  {title} {\bibinfo {title} {{Magnetic structures in the MnSb-CrSb and CrTe-CrSb systems}},\ }in\ \href@noop {} {\emph {\bibinfo {booktitle} {Proceedings of the International Conference on Magnetism}}}\ (\bibinfo  {publisher} {Institute of Physics and the Physical Society},\ \bibinfo {address} {London},\ \bibinfo {year} {1965})\ pp.\ \bibinfo {pages} {291--294}\BibitemShut {NoStop}%
\bibitem [{\citenamefont {Takei}\ \emph {et~al.}(1966)\citenamefont {Takei}, \citenamefont {Cox},\ and\ \citenamefont {Shirane}}]{take66}%
  \BibitemOpen
  \bibfield  {author} {\bibinfo {author} {\bibfnamefont {W.~J.}\ \bibnamefont {Takei}}, \bibinfo {author} {\bibfnamefont {D.~E.}\ \bibnamefont {Cox}},\ and\ \bibinfo {author} {\bibfnamefont {G.}~\bibnamefont {Shirane}},\ }\bibfield  {title} {\bibinfo {title} {{Magnetic Structures in CrTe--CrSb Solid Solutions}},\ }\href {https://doi.org/10.1063/1.1708545} {\bibfield  {journal} {\bibinfo  {journal} {J. Appl. Phys.}\ }\textbf {\bibinfo {volume} {37}},\ \bibinfo {pages} {973} (\bibinfo {year} {1966})}\BibitemShut {NoStop}%
\bibitem [{\citenamefont {Clark}\ and\ \citenamefont {Reid}(1995)}]{clar95}%
  \BibitemOpen
  \bibfield  {author} {\bibinfo {author} {\bibfnamefont {R.~C.}\ \bibnamefont {Clark}}\ and\ \bibinfo {author} {\bibfnamefont {J.~S.}\ \bibnamefont {Reid}},\ }\bibfield  {title} {\bibinfo {title} {{The analytical calculation of absorption in multifaceted crystals}},\ }\href {https://doi.org/10.1107/S0108767395007367} {\bibfield  {journal} {\bibinfo  {journal} {Acta Cryst. A}\ }\textbf {\bibinfo {volume} {51}},\ \bibinfo {pages} {887} (\bibinfo {year} {1995})}\BibitemShut {NoStop}%
\bibitem [{\citenamefont {Toby}\ and\ \citenamefont {Von~Dreele}(2013)}]{Toby_GSAS_2013}%
  \BibitemOpen
  \bibfield  {author} {\bibinfo {author} {\bibfnamefont {B.~H.}\ \bibnamefont {Toby}}\ and\ \bibinfo {author} {\bibfnamefont {R.~B.}\ \bibnamefont {Von~Dreele}},\ }\bibfield  {title} {\bibinfo {title} {{{\it GSAS-II}: the genesis of a modern open-source all purpose crystallography software package}},\ }\href {https://doi.org/10.1107/S0021889813003531} {\bibfield  {journal} {\bibinfo  {journal} {J. Appl. Cryst.}\ }\textbf {\bibinfo {volume} {46}},\ \bibinfo {pages} {544} (\bibinfo {year} {2013})}\BibitemShut {NoStop}%
\bibitem [{\citenamefont {Li}\ \emph {et~al.}(2022)\citenamefont {Li}, \citenamefont {Liu}, \citenamefont {Jiang}, \citenamefont {Jin}, \citenamefont {Pei}, \citenamefont {Wen}, \citenamefont {Yue},\ and\ \citenamefont {Wang}}]{li22}%
  \BibitemOpen
  \bibfield  {author} {\bibinfo {author} {\bibfnamefont {C.}~\bibnamefont {Li}}, \bibinfo {author} {\bibfnamefont {K.}~\bibnamefont {Liu}}, \bibinfo {author} {\bibfnamefont {D.}~\bibnamefont {Jiang}}, \bibinfo {author} {\bibfnamefont {C.}~\bibnamefont {Jin}}, \bibinfo {author} {\bibfnamefont {T.}~\bibnamefont {Pei}}, \bibinfo {author} {\bibfnamefont {T.}~\bibnamefont {Wen}}, \bibinfo {author} {\bibfnamefont {B.}~\bibnamefont {Yue}},\ and\ \bibinfo {author} {\bibfnamefont {Y.}~\bibnamefont {Wang}},\ }\bibfield  {title} {\bibinfo {title} {{Diverse Thermal Expansion Behaviors in Ferromagnetic Cr$_{1-\delta}$Te with NiAs-Type, Defective Structures}},\ }\href {https://doi.org/10.1021/acs.inorgchem.2c01826} {\bibfield  {journal} {\bibinfo  {journal} {Inorg. Chem.}\ }\textbf {\bibinfo {volume} {61}},\ \bibinfo {pages} {14641} (\bibinfo {year} {2022})}\BibitemShut {NoStop}%
\bibitem [{\citenamefont {Purwar}\ \emph {et~al.}(2023)\citenamefont {Purwar}, \citenamefont {Low}, \citenamefont {Bose}, \citenamefont {Narayan},\ and\ \citenamefont {Thirupathaiah}}]{purw23}%
  \BibitemOpen
  \bibfield  {author} {\bibinfo {author} {\bibfnamefont {S.}~\bibnamefont {Purwar}}, \bibinfo {author} {\bibfnamefont {A.}~\bibnamefont {Low}}, \bibinfo {author} {\bibfnamefont {A.}~\bibnamefont {Bose}}, \bibinfo {author} {\bibfnamefont {A.}~\bibnamefont {Narayan}},\ and\ \bibinfo {author} {\bibfnamefont {S.}~\bibnamefont {Thirupathaiah}},\ }\bibfield  {title} {\bibinfo {title} {{Investigation of the anomalous and topological Hall effects in layered monoclinic ferromagnet ${\mathrm{Cr}}_{2.76}{\mathrm{Te}}_{4}$}},\ }\href {https://doi.org/10.1103/PhysRevMaterials.7.094204} {\bibfield  {journal} {\bibinfo  {journal} {Phys. Rev. Mater.}\ }\textbf {\bibinfo {volume} {7}},\ \bibinfo {pages} {094204} (\bibinfo {year} {2023})}\BibitemShut {NoStop}%
\bibitem [{\citenamefont {Kubota}\ \emph {et~al.}(2023)\citenamefont {Kubota}, \citenamefont {Okamoto}, \citenamefont {Kanematsu}, \citenamefont {Yajima}, \citenamefont {Hirai},\ and\ \citenamefont {Takenaka}}]{kubo23}%
  \BibitemOpen
  \bibfield  {author} {\bibinfo {author} {\bibfnamefont {Y.}~\bibnamefont {Kubota}}, \bibinfo {author} {\bibfnamefont {Y.}~\bibnamefont {Okamoto}}, \bibinfo {author} {\bibfnamefont {T.}~\bibnamefont {Kanematsu}}, \bibinfo {author} {\bibfnamefont {T.}~\bibnamefont {Yajima}}, \bibinfo {author} {\bibfnamefont {D.}~\bibnamefont {Hirai}},\ and\ \bibinfo {author} {\bibfnamefont {K.}~\bibnamefont {Takenaka}},\ }\bibfield  {title} {\bibinfo {title} {{Large magnetic-field-induced strains in sintered chromium tellurides}},\ }\href {https://doi.org/10.1063/5.0134911} {\bibfield  {journal} {\bibinfo  {journal} {Appl. Phys. Lett.}\ }\textbf {\bibinfo {volume} {122}},\ \bibinfo {pages} {042404} (\bibinfo {year} {2023})}\BibitemShut {NoStop}%
\bibitem [{\citenamefont {Hatakeyama}\ \emph {et~al.}(1990)\citenamefont {Hatakeyama}, \citenamefont {Kaneko}, \citenamefont {Yoshida}, \citenamefont {Ohta},\ and\ \citenamefont {Anzai}}]{hata90}%
  \BibitemOpen
  \bibfield  {author} {\bibinfo {author} {\bibfnamefont {K.}~\bibnamefont {Hatakeyama}}, \bibinfo {author} {\bibfnamefont {T.}~\bibnamefont {Kaneko}}, \bibinfo {author} {\bibfnamefont {H.}~\bibnamefont {Yoshida}}, \bibinfo {author} {\bibfnamefont {S.}~\bibnamefont {Ohta}},\ and\ \bibinfo {author} {\bibfnamefont {S.}~\bibnamefont {Anzai}},\ }\bibfield  {title} {\bibinfo {title} {{Pressure effect on the Curie temperatures of Cr$_{1-\delta}$Te compounds}},\ }\href {https://doi.org/https://doi.org/10.1016/S0304-8853(10)80060-5} {\bibfield  {journal} {\bibinfo  {journal} {J. Magn. Magn. Mater.}\ }\textbf {\bibinfo {volume} {90--91}},\ \bibinfo {pages} {175} (\bibinfo {year} {1990})}\BibitemShut {NoStop}%
\bibitem [{\citenamefont {Fujisawa}\ \emph {et~al.}(2023)\citenamefont {Fujisawa}, \citenamefont {Pardo-Almanza}, \citenamefont {Hsu}, \citenamefont {Mohamed}, \citenamefont {Yamagami}, \citenamefont {Krishnadas}, \citenamefont {Chang}, \citenamefont {Chuang}, \citenamefont {Khoo}, \citenamefont {Zang}, \citenamefont {Soumyanarayanan},\ and\ \citenamefont {Okada}}]{fuji23}%
  \BibitemOpen
  \bibfield  {author} {\bibinfo {author} {\bibfnamefont {Y.}~\bibnamefont {Fujisawa}}, \bibinfo {author} {\bibfnamefont {M.}~\bibnamefont {Pardo-Almanza}}, \bibinfo {author} {\bibfnamefont {C.-H.}\ \bibnamefont {Hsu}}, \bibinfo {author} {\bibfnamefont {A.}~\bibnamefont {Mohamed}}, \bibinfo {author} {\bibfnamefont {K.}~\bibnamefont {Yamagami}}, \bibinfo {author} {\bibfnamefont {A.}~\bibnamefont {Krishnadas}}, \bibinfo {author} {\bibfnamefont {G.}~\bibnamefont {Chang}}, \bibinfo {author} {\bibfnamefont {F.-C.}\ \bibnamefont {Chuang}}, \bibinfo {author} {\bibfnamefont {K.~H.}\ \bibnamefont {Khoo}}, \bibinfo {author} {\bibfnamefont {J.}~\bibnamefont {Zang}}, \bibinfo {author} {\bibfnamefont {A.}~\bibnamefont {Soumyanarayanan}},\ and\ \bibinfo {author} {\bibfnamefont {Y.}~\bibnamefont {Okada}},\ }\bibfield  {title} {\bibinfo {title} {{Widely Tunable Berry Curvature in the Magnetic Semimetal Cr$_{1+\delta}$Te$_2$}},\ }\href {https://doi.org/https://doi.org/10.1002/adma.202207121} {\bibfield  {journal} {\bibinfo
  {journal} {Adv. Mater.}\ }\textbf {\bibinfo {volume} {35}},\ \bibinfo {pages} {2207121} (\bibinfo {year} {2023})}\BibitemShut {NoStop}%
\bibitem [{\citenamefont {Dijkstra}\ \emph {et~al.}(1989)\citenamefont {Dijkstra}, \citenamefont {Weitering}, \citenamefont {van Bruggen}, \citenamefont {Haas},\ and\ \citenamefont {de~Groot}}]{dijk89}%
  \BibitemOpen
  \bibfield  {author} {\bibinfo {author} {\bibfnamefont {J.}~\bibnamefont {Dijkstra}}, \bibinfo {author} {\bibfnamefont {H.~H.}\ \bibnamefont {Weitering}}, \bibinfo {author} {\bibfnamefont {C.~F.}\ \bibnamefont {van Bruggen}}, \bibinfo {author} {\bibfnamefont {C.}~\bibnamefont {Haas}},\ and\ \bibinfo {author} {\bibfnamefont {R.~A.}\ \bibnamefont {de~Groot}},\ }\bibfield  {title} {\bibinfo {title} {{Band-structure calculations, and magnetic and transport properties of ferromagnetic chromium tellurides (CrTe, Cr$_3$Te$_4$, Cr$_2$Te$_3$)}},\ }\href {https://doi.org/10.1088/0953-8984/1/46/008} {\bibfield  {journal} {\bibinfo  {journal} {J. Phys. Condens. Matter}\ }\textbf {\bibinfo {volume} {1}},\ \bibinfo {pages} {9141} (\bibinfo {year} {1989})}\BibitemShut {NoStop}%
\bibitem [{\citenamefont {Bose}\ \emph {et~al.}(2025)\citenamefont {Bose}, \citenamefont {Purwar}, \citenamefont {Thirupathaiah},\ and\ \citenamefont {Narayan}}]{bose25}%
  \BibitemOpen
  \bibfield  {author} {\bibinfo {author} {\bibfnamefont {A.}~\bibnamefont {Bose}}, \bibinfo {author} {\bibfnamefont {S.}~\bibnamefont {Purwar}}, \bibinfo {author} {\bibfnamefont {S.}~\bibnamefont {Thirupathaiah}},\ and\ \bibinfo {author} {\bibfnamefont {A.}~\bibnamefont {Narayan}},\ }\bibfield  {title} {\bibinfo {title} {{Anomalous and parallel Hall effects in ferromagnetic Weyl metal ${\mathrm{Cr}}_{3}{\mathrm{Te}}_{4}$}},\ }\href {https://doi.org/10.1103/PhysRevMaterials.9.044413} {\bibfield  {journal} {\bibinfo  {journal} {Phys. Rev. Mater.}\ }\textbf {\bibinfo {volume} {9}},\ \bibinfo {pages} {044413} (\bibinfo {year} {2025})}\BibitemShut {NoStop}%
\bibitem [{\citenamefont {Huang}\ \emph {et~al.}(2025)\citenamefont {Huang}, \citenamefont {Zuo}, \citenamefont {Zhang}, \citenamefont {Xing}, \citenamefont {Yao}, \citenamefont {Zhang}, \citenamefont {Ma}, \citenamefont {Xu}, \citenamefont {Jiao}, \citenamefont {Zhou}, \citenamefont {Sankar}, \citenamefont {Qian},\ and\ \citenamefont {Xu}}]{huan25}%
  \BibitemOpen
  \bibfield  {author} {\bibinfo {author} {\bibfnamefont {Y.}~\bibnamefont {Huang}}, \bibinfo {author} {\bibfnamefont {N.}~\bibnamefont {Zuo}}, \bibinfo {author} {\bibfnamefont {Z.}~\bibnamefont {Zhang}}, \bibinfo {author} {\bibfnamefont {X.}~\bibnamefont {Xing}}, \bibinfo {author} {\bibfnamefont {X.}~\bibnamefont {Yao}}, \bibinfo {author} {\bibfnamefont {A.}~\bibnamefont {Zhang}}, \bibinfo {author} {\bibfnamefont {H.}~\bibnamefont {Ma}}, \bibinfo {author} {\bibfnamefont {C.}~\bibnamefont {Xu}}, \bibinfo {author} {\bibfnamefont {W.}~\bibnamefont {Jiao}}, \bibinfo {author} {\bibfnamefont {W.}~\bibnamefont {Zhou}}, \bibinfo {author} {\bibfnamefont {R.}~\bibnamefont {Sankar}}, \bibinfo {author} {\bibfnamefont {D.}~\bibnamefont {Qian}},\ and\ \bibinfo {author} {\bibfnamefont {X.}~\bibnamefont {Xu}},\ }\bibfield  {title} {\bibinfo {title} {{In-Plane Magnetic Anisotropy and Large Topological Hall Effect in Self-Intercalated Ferromagnet Cr$_{1.61}$Te$_2$}},\ }\href
  {https://doi.org/https://doi.org/10.1002/adfm.202510351} {\bibfield  {journal} {\bibinfo  {journal} {Adv. Funct. Mater.}\ }\textbf {\bibinfo {volume} {n/a}},\ \bibinfo {pages} {e10351} (\bibinfo {year} {2025})}\BibitemShut {NoStop}%
\bibitem [{\citenamefont {Saha}\ \emph {et~al.}(2022)\citenamefont {Saha}, \citenamefont {Meyerheim}, \citenamefont {G\"obel}, \citenamefont {Hazra}, \citenamefont {Deniz}, \citenamefont {Mohseni}, \citenamefont {Antonov}, \citenamefont {Ernst}, \citenamefont {Knyazev}, \citenamefont {Bedoya-Pinto}, \citenamefont {Mertig},\ and\ \citenamefont {Parkin}}]{saha22}%
  \BibitemOpen
  \bibfield  {author} {\bibinfo {author} {\bibfnamefont {R.}~\bibnamefont {Saha}}, \bibinfo {author} {\bibfnamefont {H.~L.}\ \bibnamefont {Meyerheim}}, \bibinfo {author} {\bibfnamefont {B.}~\bibnamefont {G\"obel}}, \bibinfo {author} {\bibfnamefont {B.~K.}\ \bibnamefont {Hazra}}, \bibinfo {author} {\bibfnamefont {H.}~\bibnamefont {Deniz}}, \bibinfo {author} {\bibfnamefont {K.}~\bibnamefont {Mohseni}}, \bibinfo {author} {\bibfnamefont {V.}~\bibnamefont {Antonov}}, \bibinfo {author} {\bibfnamefont {A.}~\bibnamefont {Ernst}}, \bibinfo {author} {\bibfnamefont {D.}~\bibnamefont {Knyazev}}, \bibinfo {author} {\bibfnamefont {A.}~\bibnamefont {Bedoya-Pinto}}, \bibinfo {author} {\bibfnamefont {I.}~\bibnamefont {Mertig}},\ and\ \bibinfo {author} {\bibfnamefont {S.~S.~P.}\ \bibnamefont {Parkin}},\ }\bibfield  {title} {\bibinfo {title} {{Observation of N\'eel-type skyrmions in acentric self-intercalated Cr$_{1+\delta}$Te$_2$}},\ }\href {https://doi.org/10.1038/s41467-022-31319-y} {\bibfield  {journal} {\bibinfo  {journal}
  {Nat. Commun.}\ }\textbf {\bibinfo {volume} {13}},\ \bibinfo {pages} {3965} (\bibinfo {year} {2022})}\BibitemShut {NoStop}%
\bibitem [{\citenamefont {Pradhan}\ \emph {et~al.}(2024)\citenamefont {Pradhan}, \citenamefont {Liu}, \citenamefont {Zheng}, \citenamefont {Song},\ and\ \citenamefont {Wu}}]{prad24}%
  \BibitemOpen
  \bibfield  {author} {\bibinfo {author} {\bibfnamefont {S.~K.}\ \bibnamefont {Pradhan}}, \bibinfo {author} {\bibfnamefont {Y.}~\bibnamefont {Liu}}, \bibinfo {author} {\bibfnamefont {F.}~\bibnamefont {Zheng}}, \bibinfo {author} {\bibfnamefont {D.}~\bibnamefont {Song}},\ and\ \bibinfo {author} {\bibfnamefont {R.}~\bibnamefont {Wu}},\ }\bibfield  {title} {\bibinfo {title} {{Observation of N\'eel-type magnetic skyrmion in a layered non-centrosymmetric itinerant ferromagnet CrTe$_{1.38}$}},\ }\href {https://doi.org/10.1063/5.0231254} {\bibfield  {journal} {\bibinfo  {journal} {Appl. Phys. Lett.}\ }\textbf {\bibinfo {volume} {125}},\ \bibinfo {pages} {152402} (\bibinfo {year} {2024})}\BibitemShut {NoStop}%
\bibitem [{\citenamefont {Rai}\ \emph {et~al.}(2025)\citenamefont {Rai}, \citenamefont {Kuila}, \citenamefont {Saha}, \citenamefont {Hazra}, \citenamefont {De}, \citenamefont {Sau}, \citenamefont {Gopalan}, \citenamefont {Jana}, \citenamefont {Parkin},\ and\ \citenamefont {Kumar}}]{rai25}%
  \BibitemOpen
  \bibfield  {author} {\bibinfo {author} {\bibfnamefont {B.}~\bibnamefont {Rai}}, \bibinfo {author} {\bibfnamefont {S.~K.}\ \bibnamefont {Kuila}}, \bibinfo {author} {\bibfnamefont {R.}~\bibnamefont {Saha}}, \bibinfo {author} {\bibfnamefont {S.}~\bibnamefont {Hazra}}, \bibinfo {author} {\bibfnamefont {C.}~\bibnamefont {De}}, \bibinfo {author} {\bibfnamefont {J.}~\bibnamefont {Sau}}, \bibinfo {author} {\bibfnamefont {V.}~\bibnamefont {Gopalan}}, \bibinfo {author} {\bibfnamefont {P.~P.}\ \bibnamefont {Jana}}, \bibinfo {author} {\bibfnamefont {S.~S.~P.}\ \bibnamefont {Parkin}},\ and\ \bibinfo {author} {\bibfnamefont {N.}~\bibnamefont {Kumar}},\ }\bibfield  {title} {\bibinfo {title} {{Peculiar Magnetic and Magneto-Transport Properties in a Noncentrosymmetric Self-Intercalated van der Waals Ferromagnet Cr$_5$Te$_8$}},\ }\href {https://doi.org/10.1021/acs.chemmater.4c02996} {\bibfield  {journal} {\bibinfo  {journal} {Chem. Mater.}\ }\textbf {\bibinfo {volume} {37}},\ \bibinfo {pages} {746} (\bibinfo {year}
  {2025})}\BibitemShut {NoStop}%
\bibitem [{\citenamefont {Zhang}\ \emph {et~al.}(2022)\citenamefont {Zhang}, \citenamefont {Liu}, \citenamefont {Niu}, \citenamefont {Lu}, \citenamefont {Wang}, \citenamefont {Sarikhani}, \citenamefont {Wu}, \citenamefont {Zhu}, \citenamefont {Sun}, \citenamefont {Vaninger}, \citenamefont {Miceli}, \citenamefont {Li}, \citenamefont {Singh}, \citenamefont {Hor}, \citenamefont {Zhao}, \citenamefont {Liu}, \citenamefont {He}, \citenamefont {Zhang}, \citenamefont {Bian}, \citenamefont {Yu},\ and\ \citenamefont {Xu}}]{zhan22}%
  \BibitemOpen
  \bibfield  {author} {\bibinfo {author} {\bibfnamefont {X.}~\bibnamefont {Zhang}}, \bibinfo {author} {\bibfnamefont {W.}~\bibnamefont {Liu}}, \bibinfo {author} {\bibfnamefont {W.}~\bibnamefont {Niu}}, \bibinfo {author} {\bibfnamefont {Q.}~\bibnamefont {Lu}}, \bibinfo {author} {\bibfnamefont {W.}~\bibnamefont {Wang}}, \bibinfo {author} {\bibfnamefont {A.}~\bibnamefont {Sarikhani}}, \bibinfo {author} {\bibfnamefont {X.}~\bibnamefont {Wu}}, \bibinfo {author} {\bibfnamefont {C.}~\bibnamefont {Zhu}}, \bibinfo {author} {\bibfnamefont {J.}~\bibnamefont {Sun}}, \bibinfo {author} {\bibfnamefont {M.}~\bibnamefont {Vaninger}}, \bibinfo {author} {\bibfnamefont {P.~F.}\ \bibnamefont {Miceli}}, \bibinfo {author} {\bibfnamefont {J.}~\bibnamefont {Li}}, \bibinfo {author} {\bibfnamefont {D.~J.}\ \bibnamefont {Singh}}, \bibinfo {author} {\bibfnamefont {Y.~S.}\ \bibnamefont {Hor}}, \bibinfo {author} {\bibfnamefont {Y.}~\bibnamefont {Zhao}}, \bibinfo {author} {\bibfnamefont {C.}~\bibnamefont {Liu}}, \bibinfo {author}
  {\bibfnamefont {L.}~\bibnamefont {He}}, \bibinfo {author} {\bibfnamefont {R.}~\bibnamefont {Zhang}}, \bibinfo {author} {\bibfnamefont {G.}~\bibnamefont {Bian}}, \bibinfo {author} {\bibfnamefont {D.}~\bibnamefont {Yu}},\ and\ \bibinfo {author} {\bibfnamefont {Y.}~\bibnamefont {Xu}},\ }\bibfield  {title} {\bibinfo {title} {{Self-Intercalation Tunable Interlayer Exchange Coupling in a Synthetic van der Waals Antiferromagnet}},\ }\href {https://doi.org/https://doi.org/10.1002/adfm.202202977} {\bibfield  {journal} {\bibinfo  {journal} {Adv. Funct. Mater.}\ }\textbf {\bibinfo {volume} {32}},\ \bibinfo {pages} {2202977} (\bibinfo {year} {2022})}\BibitemShut {NoStop}%
\bibitem [{\citenamefont {Conner}\ \emph {et~al.}(2024)\citenamefont {Conner}, \citenamefont {Sarikhani}, \citenamefont {Volz}, \citenamefont {Vaninger}, \citenamefont {He}, \citenamefont {Kelley}, \citenamefont {Cook}, \citenamefont {Sah}, \citenamefont {Clark}, \citenamefont {Lucker}, \citenamefont {Zhang}, \citenamefont {Miceli}, \citenamefont {Hor}, \citenamefont {Zhang},\ and\ \citenamefont {Bian}}]{conn24}%
  \BibitemOpen
  \bibfield  {author} {\bibinfo {author} {\bibfnamefont {C.}~\bibnamefont {Conner}}, \bibinfo {author} {\bibfnamefont {A.}~\bibnamefont {Sarikhani}}, \bibinfo {author} {\bibfnamefont {T.}~\bibnamefont {Volz}}, \bibinfo {author} {\bibfnamefont {M.}~\bibnamefont {Vaninger}}, \bibinfo {author} {\bibfnamefont {X.}~\bibnamefont {He}}, \bibinfo {author} {\bibfnamefont {S.}~\bibnamefont {Kelley}}, \bibinfo {author} {\bibfnamefont {J.}~\bibnamefont {Cook}}, \bibinfo {author} {\bibfnamefont {A.}~\bibnamefont {Sah}}, \bibinfo {author} {\bibfnamefont {J.}~\bibnamefont {Clark}}, \bibinfo {author} {\bibfnamefont {H.}~\bibnamefont {Lucker}}, \bibinfo {author} {\bibfnamefont {C.}~\bibnamefont {Zhang}}, \bibinfo {author} {\bibfnamefont {P.}~\bibnamefont {Miceli}}, \bibinfo {author} {\bibfnamefont {Y.~S.}\ \bibnamefont {Hor}}, \bibinfo {author} {\bibfnamefont {X.}~\bibnamefont {Zhang}},\ and\ \bibinfo {author} {\bibfnamefont {G.}~\bibnamefont {Bian}},\ }\href {https://arxiv.org/abs/2411.13721} {\bibinfo {title} {{Enhanced
  Antiferromagnetic Phase in Metastable Self-Intercalated Cr$_{1+x}$Te$_2$ Compounds}}} (\bibinfo {year} {2024}),\ \Eprint {https://arxiv.org/abs/2411.13721} {arXiv:2411.13721} \BibitemShut {NoStop}%
\end{thebibliography}%

\end{document}